%% file: sample-manuscript.tex
\documentclass[manuscript, acmlarge]{acmart}

\AtBeginDocument{%
  \providecommand\BibTeX{{%
    Bib\TeX}}}

\setcopyright{acmlicensed}
\copyrightyear{2018}
\acmYear{2018}
\acmDOI{XXXXXXX.XXXXXXX}
\acmConference[Conference acronym 'XX]{Make sure to enter the correct
  conference title from your rights confirmation email}{June 03--05,
  2018}{Woodstock, NY}
\acmISBN{978-1-4503-XXXX-X/2018/06}

\usepackage{textgreek}
\usepackage{algpseudocode}
\usepackage[linesnumbered,titlenumbered,ruled,noend]{algorithm2e}
\usepackage{algorithmicx}
\usepackage{subcaption}

\acmYear{2026}

\begin{document}

\title{E-MagDiP: \b{E}lectro\b{-}\b{M}\b{a}\b{g}netic based \b{D}\b{i}fferential \b{P}rivacy for EEG based Community Sensing}


\author{Ayanga Imesha Kumari Kalupahana}
\affiliation{%
  \institution{National University of Singapore}
  \country{Singapore}
 }
\email{ayangaim@comp.nus.edu.sg}
\orcid{0009-0005-5653-9165}

\author{Vishruti Ranjan}
\affiliation{%
  \institution{National University of Singapore}
  \country{Singapore}
}
\email{vishruti@comp.nus.edu.sg}
\orcid{0009-0006-4488-548X}

\author{Li-Shiuan Peh}
\affiliation{%
  \institution{National University of Singapore}
  \country{Singapore}
 }
\email{peh@nus.edu.sg}
\orcid{0000-0001-9010-6519}

\renewcommand{\shortauthors}{Kalupahana et al.}
\input{sections/abstract}




\maketitle
\input{sections/introduction}

\input{sections/Threat_Model}
\input{sections/Differential_Privacy_for_EEG}

\input{sections/Related_work}

\input{sections/System_Design}

\input{sections/Experimental_Setup}

\input{sections/Evaluation}

\input{sections/Discussion}

\input{sections/limitations}
\input{sections/conclusion}
\bibliographystyle{ACM-Reference-Format}
\bibliography{reference}
\appendix

\input{sections/Serandip_applicability}

\end{document}

%% file: sections/abstract.tex
\begin{abstract}
EEG-based community sensing programs are emerging globally as a tool to leverage aggregated brain data to gain insights into attentiveness of students and employees\footnote{Brain-reading technology : \label{employee_eeg}\href{https://u.osu.edu/mclc/2018/05/01/brain-reading-technology/}{https://u.osu.edu/mclc/2018/05/01/brain-reading-technology/}}
\footnote{How China Is Using Artificial Intelligence in Classrooms | WSJ : \label{classroom_eeg}\href{https://www.youtube.com/watch?v=JMLsHI8aV0g}{https://www.youtube.com/watch?v=JMLsHI8aV0g}}. 
But these programs raise privacy concerns because EEG signals contain sensitive personal information. Differential Privacy (DP) can protect individuals while preserving aggregate statistics, yet applying DP to EEG data is challenging as it requires user-level noise generation, which increases power and latency. Besides, most commercial EEG headsets cannot be modified to add such noise. We propose E-MagDiP, a framework that uses an external radio to transmit RF signals onto EEG headsets, perturbing signals at acquisition to induce DP noise. Experiments on three off-the-shelf headsets show 
E-MagDiP's ability of providing DP guarantee 38.12 for 100 participants. To the best of our knowledge, E-MagDiP is the first framework to use RF signals for privacy instead of attacks, enabling practical DP for EEG community sensing without any user-level modification.

\end{abstract}

%% file: sections/introduction.tex
\section{INTRODUCTION}
Electroencephalogram (EEG) systems measure electrical activity of the brain through electrodes on the scalp and are widely used in applications like Brain Computer Interfaces (BCIs) and mental-state sensing~\cite{Mobisys_EEG_applications}. More recently, the rise of portable consumer-grade devices has been leveraged to measure brain function in community settings \cite{community_eeg}. This has been applied to variety of scenarios, like assessing factory productivity improvement under new machinery~\cite{halt_of_eeg_productionline_employees}, evaluating the efficacy of new teaching methods~\cite{China_eeg_classroom,halt_of_China_eeg_classroom, headphone_eeg}, and studying driver drowsiness for improving vehicle safety features and road conditions~\cite{nature1EEGFatigue,driver_fatigue}. These applications illustrate the benefits of EEG-based community sensing. 

EEG data is sensitive because it contains deep insights into processes that contribute to cognitive, emotional and instinctual functions. EEG has previously demonstrated the ability to predict personal traits, and even uncover information like user's PIN numbers, bank account details, romantic attractions, and skill levels in various tasks \cite{MAGEE20243017, martinovic2012}. If raw EEG data falls into the wrong hands, it could adversely affect those whose data has been leaked. In community sensing settings, raw EEG data is often exposed to a data aggregator and stored for further analytics~\cite{eeg_anonimazation_not_enough}. These aggregators hold considerable responsibility for protecting these data; despite this, many institutions have failed to do so. For example, the LAUSD \cite{LAUSD} breach in 2022 allowed a hacking group to leak students' personal, health, and psychological information online. Similarly, in 2018, a breach in a state-owned Aadhar~\cite{databreaches} system exposed photos, fingerprints, and retinal scans of over a billion Indian citizens. Such instances, along with the lax industry practices in storing cognitive data \cite{unesco_neuro} create fear and apprehension in individuals from participating in community sensing activities. This has led to some EEG-based community sensing programs being halted~\cite{halt_of_China_eeg_classroom}.

  EEG signals are considered biometric signatures as they possess unique features tied to individuals~\cite{eeg_biometric,Mobisys_eeg_as_passwords}. Thus, standard mechanisms like anonymization fail in community sensing~\cite{anonymization_not_enough2,eeg_anonimazation_not_enough}.
  Mature techniques like secure-pairing and encryption  are able to authorize access to data, but preservation of privacy with untrusted community sensing aggregators cannot be simply resolved through authorization.

To address this, Differential Privacy (DP) has been proposed~\cite{DP_in_CHI,DP_CHI_2,DP_CHI_3} to realise the potential of EEG community sensing programs~\cite{Ienca_Fins_Jox_Jotterand_Voeneky_Andorno_Ball_Castelluccia_Chavarriaga_Chneiweiss_et, Yuste_Goering_Arcas_Bi_Carmena_Carter_Fins_Friesen_Gallant_Huggins_et}. DP can hide an individual's presence by perturbing their data with statistical noise. One way is to add this noise to the data after collection. But, there have been attacks conducted whilst the data is being transmitted (like this proximity attack on Neurosky \cite{EEG_security_attack}). Additionally, the rising prevalence of WiFi-based communication of EEG headsets with servers~\cite{cyton_16kHz,aireeg,neurotrace} exacerbates the need for implementing statistical noise generation on the headsets to realize DP. However, injecting DP on the headsets is not practical. Most commercial EEG headsets are closed systems, they leave no access to modify their onboard data pipelines. Moreover, even if manufacturers choose to support this, wearable devices face tight power and computation limits \cite{serandip}, adding per sample perturbation can further constrain their already limited sampling rates.

So, is there a way to provide DP without modifying existing EEG hardware? We draw inspiration from an unexpected source: prior work on EM-based attacks against analog sensors~\cite{Ghost_Talk, EM_attack_against_embeded_system, Selvaraj2018IntentionalEI} and EEG systems~\cite{ brain_hack, first_to_EEG_attack}. These attacks use amplitude-modulated (AM) RF signals to deliberately corrupt sensor readings. 
Instead, we propose \textbf{\textit{E-MagDiP }}which leverages electromagnetic (EM) signals for defence instead of attack. E-MagDiP performs AM on a white Gaussian signal and transmits it as an RF signal. When targeted electrodes or leads are exposed to the RF signal, it induces controlled Gaussian noise on the EEG stream, thereby achieving differential privacy without any modification to the EEG headset's underlying hardware/software.

E-MagDiP has been prototyped using the USRP B200 software-defined radio (SDR) interfaced with a PC. To evaluate our concept, we use three off-the-shelf EEG headsets: research grade OpenBCI Cyton, consumer grade Neurosky MindWave Mobile 2 and Sichiray Taurus 2.0 Brainwave, and show that E-MagDiP provides a DP guarantee of ($\epsilon$) of 38.12 to 100 participants across all three unmodified commercial devices.
Statistical noise addition to data can potentially reduce classification accuracy; however, E-MagDiP delivers improved accuracy of 1.2\% over a state-of-the-art software-based DP method. 
We summarize our contributions as follows: 
\begin{itemize}
    \item We present \textbf{\textit{E-MagDiP}}, which, to the best of our knowledge is the first method to provide DP for off-the-shelf EEG headsets for community sensing.
    \item We introduce a mechanism that provides required DP without any modifications to the EEG headsets.
    \item This is achieved by design of an RF-based perturbation technique that induces calibrated white Gaussian noise on EEG signals via AM transmission.
    \item We validate E-MagDiP on three commercial EEG systems: OpenBCI Cyton, Neurosky MindWave Mobile 2, and Sichiray Taurus 2.0 Brainwave.
    \item In user trials with 10 participants, E-MagDiP achieves DP with minimal impact on classification accuracy.
\end{itemize}

The remaining of the paper is structured as follows. In Section 2, we discuss deployment scenario. Section 3 introduces background on DP theory, Gaussianity of amplitude modulated radio frequency signals and its impact on sensors, wires and electrodes, impact and quantification of white Gaussian noise induced on EEG and DP noise requirement in EEG based community sensing. In Section 4, we review  the literature review on differential privacy preservation techniques for wearable community sensing applications, differential privacy without random noise generation and privacy and security attacks and protections for EEG and BCI. Section 5 outlines the various steps involved in our E-MagDiP framework. Experimental setup for validation of our solution against existing DP solutions is  presented in Section 6. Section 7 presents experimental validation results while Section 8 discusses the off-the shelf EEG headsets' technical specifications which limit/benefit E-MagDiP's applicability. Health and regulatory concerns related to E-MagDiP deployment are discussed as limitations under Section 9. Finally, we conclude the paper in Section 10.

%% file: sections/Threat_Model.tex
\section{ DEPLOYMENT  SCENARIO }
\begin{figure}[htb]
    \centering
    \includegraphics[width=1\linewidth]{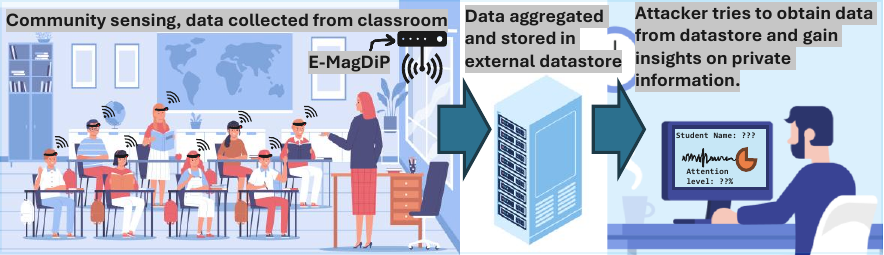}
    \caption{E-MagDiP prevents attackers who can access aggregated data from being able to identify an individual's private information and related data.}
    \label{fig:attack_scenario}
    \vspace{-1\baselineskip}
\end{figure}

Figure \ref{fig:attack_scenario} illustrates our deployment model. Consider a scenario like a classroom, where $N$ participants wear off-the-shelf EEG headsets. A trusted data collector deploys E-MagDiP \textit{inside the same room as the participants.} The transmitted EEG data is now differentially private and is stored in a remote analysis server \textit{outside} the room. However, entities with access to this aggregated dataset may be curious or untrusted. \\
\textbf{Attacker's objective:} The attacker seeks to infer private information about a specific participant $n$, like cognitive state indicators. \\
\textbf{Attacker's capability:} The attacker can observe aggregated EEG traces and has access to the data from remaining $N - 1$ participants in similar tasks. This side information aids in reconstruction and inference tasks of participant $n$~\cite{2020_Census_Disclosure_Avoidance_System,sensitive_data_reveal_fl}. \\
\textbf{Is encryption sufficient?} No. Encryption protects unauthorized access of data. DP ensures that even if the data is exposed to the authorized but untrusted aggregator, they are unable to identify individuals from the data~\cite{2020_Census_Disclosure_Avoidance_System}. 

%% file: sections/Differential_Privacy_for_EEG.tex
\section{THEORY AND BACKGROUND}
\subsection{DP in community sensing}
\label{Differential_privacy_theorems}

The key mathematical definitions and theorems of differential privacy that apply to EEG-based community sensing are summarized in this section.

\begin{definition}[Differential Privacy~\cite{dwork}]
A privacy mechanism M satisfies $(\epsilon, \delta)$-differential privacy ($\epsilon, \delta >0$), if we have
\begin{equation}
Pr[M(D_1)\in S] \leq \exp(\epsilon) \cdot Pr [M(D_2)\in S] + \delta
\end{equation}
for all adjacent EEG databases $D_1$  and  $D_2$  which differ on at most one participant's EEG record, with the participant's data present/absent from the adjacent database, for all sets S  $\subseteq$  Range(M). Typically, the Gaussian mechanism is used to implement DP where random noise is drawn from a zero-mean Gaussian distribution and added to the query output $f(D)$ with the Gaussian noise scaled to  $l_2$ sensitivity~\cite{DPbook}:
\end{definition}

\begin{definition}[ $l_2$ sensitivity~\cite{DPbook}]
For any function $f:D  \rightarrow R^d$, the $l_2$ sensitivity of $f$ w.r.t.\ $D$ is given as    
\begin{equation}
    \Delta_2(f) =\max\limits_{D_1, D_2 \in D } \|f(D_1)-f(D_2)\|_2   
\end{equation}
for all $D_1, D_2$ differing on at most one record. $\Delta_2(f)$ measures the Euclidean distance as follows:
\begin{equation}
   \Delta_2(f)=\sqrt{\sum_i[f_i(D_1)-f_i(D_2)]^2} 
\end{equation}
For any function $f:D  \rightarrow R^d$, the {\it Gaussian mechanism} with parameter $\sigma$ for any dataset $D$ is $M(D)= f(D) + N(0,\sigma^2 )^d$:
\end{definition}

\begin{theorem}[Gaussian Mechanism~\cite{DPbook}]
\label{theorem:1}
Let $\epsilon \in (0, 1]$ be arbitrary. For $c^2 > 2 \ln(1.25/\delta)$, the Gaussian mechanism with parameter $\sigma \geq c \Delta_2(f) /\epsilon$ is $(\epsilon, \delta )$-differentially private.
\end{theorem} 
In addition, based on the results in~\cite{Mironov17}, we have the following corollary on the overall privacy guarantee of multiple invocations of the Gaussian mechanism.
\begin{corollary}
Let $f_1,\ldots, f_k$ be a set of functions. For $c^2> 2 \ln(1.25/\delta)$, applying Gaussian mechanism on $f_1,\ldots, f_k$ with parameter $\sigma \geq \sqrt{k} \cdot c \Delta_2(f) /\epsilon$ is $(\epsilon, \delta)$-differentially private.
\label{corollary:1}
\end{corollary}
To understand the DP requirement for EEG based community sensing, we apply Theorem~\ref{theorem:1} to estimate the standard deviation of privacy noise required from each participant.  $\delta$ in Theorem~\ref{theorem:1}  is represented as 1/(number of participants in the EEG-based community sensing program) and $\Delta_2(f)$ is the EEG measurement range (difference between the maximum and minimum EEG sensor reading). The standard deviation estimated using Theorem~\ref{theorem:1} is the total standard deviation of accumulated privacy noise from all the participants under a typical EEG based Distributed Differential Privacy (DDP) setup $(\sigma\textsubscript{a+b+....})$. The total variance of the required privacy noise is the sum of variances across all participants: 
\begin{equation}
    \sigma^2_{a+b+...+N} = \sigma^2_a + \sigma^2_b +....+ \sigma^2_N 
\end{equation}
\label{DDP_theory}
Therefore, the standard deviation of the privacy noise required from each user to provide a $(\epsilon, \delta)-$DDP guarantee would be obtained as
\begin{equation}
\label{equation:7}
    \sigma_a = \sigma_{a+b+...+N}/\sqrt{n}
\end{equation}

Most machine learning-based classifiers used in community sensing require consecutive time-series EEG data to accurately estimate a participant’s behavior at a given time. Given the classifier requires $j$ consecutive samples from $k$ channels of the EEG sensors as input,  we know from Corollary~\ref{corollary:1} that $\sqrt{j} \cdot\sqrt{k}\cdot c \Delta_2(f) /\epsilon$  would be the required SD of white Gaussian noise in the EEG data to ensure differential privacy.  Therefore, each user's EEG data should have white Gaussian noise with standard deviation ($\sigma_a$):
\begin{equation}
\label{equation:8}
\sigma_a \geq \sqrt{j} \cdot\sqrt{k} \cdot c \Delta_2(f) /{\sqrt{n}\epsilon}   
\end{equation}
to ensure $(\epsilon, \delta)$-DDP guarantee at the server used in community sensing, after classification. Thus, the required SD of privacy noise increases proportionally to $\sqrt{jk}$. Where $j$ is the number of consecutive timeseries EEG samples and $k$ is the number of EEG channels used by the classifier.
\subsection{Homomorphic encryption with DP}

Since we are providing DDP, the data is perturbed with less noise compared to local DP~\cite{rastogi,shi}. In other words, though noise is added at the user level, privacy is guaranteed only at the aggregated level. Hence additional security measures need to be taken for DDP. This is provided via Homomorphic Encryption (HE). HE allows servers to perform computations on the user data in its encrypted format~\cite{homo_encry,emotion_he}. Hence, user-level data must first be homomorphically encrypted to generate an aggregated value in an encrypted format. 

\subsection{Gaussianity of Amplitude Modulated (AM) radio frequency signals and its impact on sensors, wires and electrodes}

     According to Sung et al.~\cite{approximation_of_phase_AM}, the envelope of Amplitude Modulated (AM) signal ($e(t)$) can be represented as:
     \begin{equation}
\label{equation:9}
        e(t)= A_c (1+m\cdot A_m(t))
\end{equation}
    where $A_C$ is carrier amplitude, $m$ is the modulation index where 0$\leq$ $m$ $\leq$1 and $A_m$(t) is the amplitude of the message signal at time $t$. When $\rho$= $A^2_c$/2$\sigma^2$ is sufficiently large, where $\sigma$ is the SD of narrowband noise, $e(t)$ follows a Gaussian distribution, so its SD is $\sqrt{a}\cdot\sigma$ where a= 2$\cdot\rho m^2 \sigma^2_m$+1 and $\sigma_m$ is modulating function's Gaussian process's SD. Thus, the SD of the Gaussian distribution following the envelope of the AM signal ($e(t)$) is:
      \begin{equation}
       \label{equation:10}
        \sigma_{e(t)}=\sqrt{A^2_c\cdot m^2\cdot \sigma^2_m +\sigma ^2}
\end{equation}
     In short, 
      the SD of AM signals can be increased by raising the amplitude of the carrier signal ($A_c$), setting modulation index $(m)$ to 1 and increasing the SD of the modulating signal ($\sigma^2_m$). These properties of AM signals have been exploited in prior attacks on analog sensors~\cite{Ghost_Talk,EM_attack_against_embeded_system, Selvaraj2018IntentionalEI} and wired electrodes~\cite{brain_hack,first_to_EEG_attack} by transmitting a targeted signal as a modulated message signal with a coupled carrier frequency($f_c$). 

    In this work, to induce white Gaussian noise on EEG headsets, we select a white Gaussian noise signal as the message signal. We then aim to maximise the amplitude of the carrier signal while maintaining a modulation index of $m = 1$. This in turn increases the SD of the Gaussian message signal, thereby inducing stronger white Gaussian noise on EEG data to meet DP noise requirements. Section ~\ref{sec:5} details how this underlying principle informs our overall system design.

\subsection{Impact and quantification of white Gaussian noise induced on EEG }    

\begin{figure} [hbt!]
            \centering
            \includegraphics[width=0.8\linewidth]{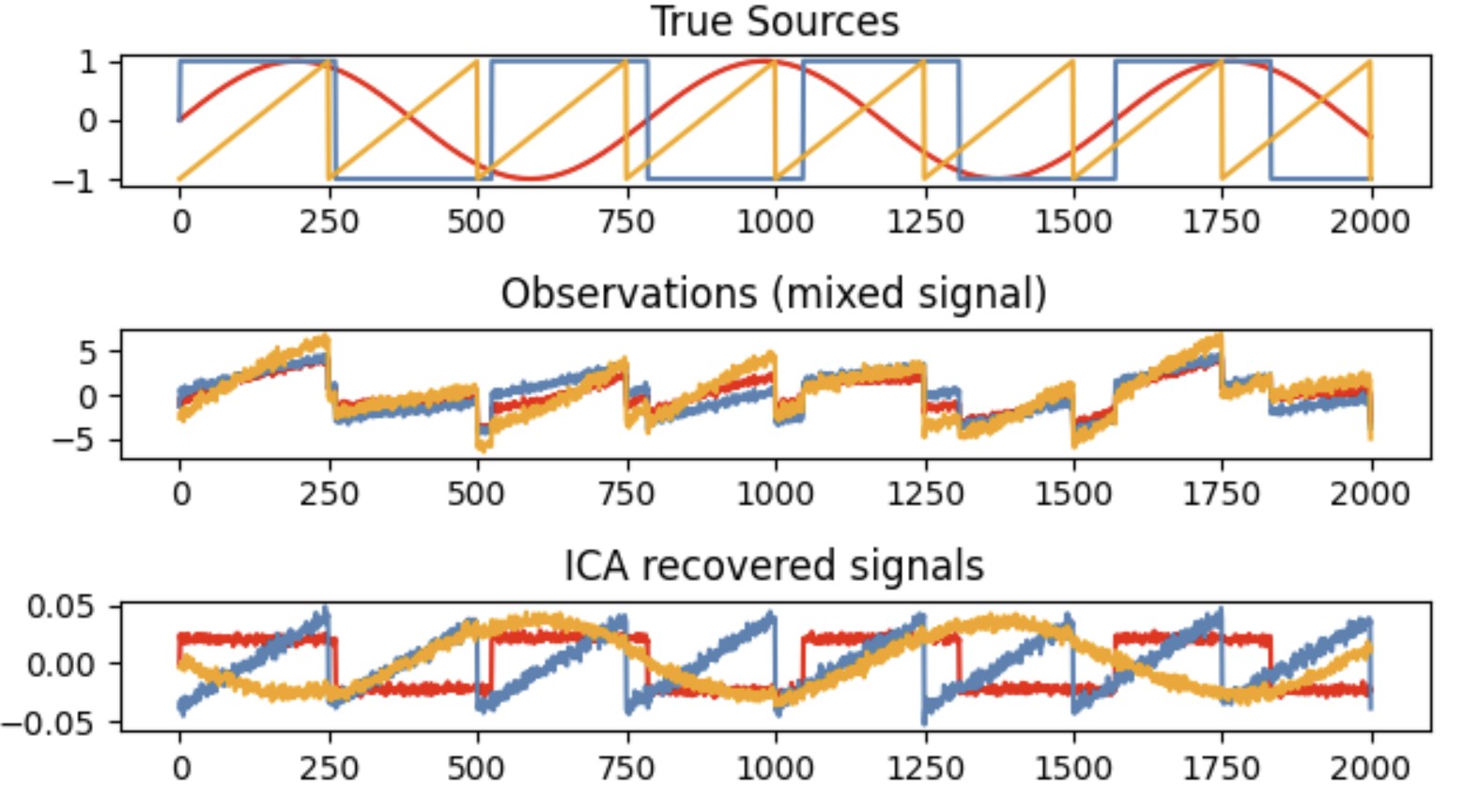}
            \caption{(a) The original signal is comprised of sources at 1.33Hz sine (red), 2Hz square (blue), and 4Hz sawtooth (yellow) waves. (b) Mixed signals wave (c) Effect of ICA applied to mixed signals wave -- added Gaussian noise (scale=0.1) did not affect ICA's recovery. Experiments were simulated using python script in Google Colab with original signal amplitude of 1.} 
            \label{fig:ica_impact}
            \vspace{-\baselineskip}
            \end{figure}

\begin{figure} [hbt!]
 \vspace{-0.5\baselineskip}
            \centering
           \includegraphics[width=0.8\linewidth]{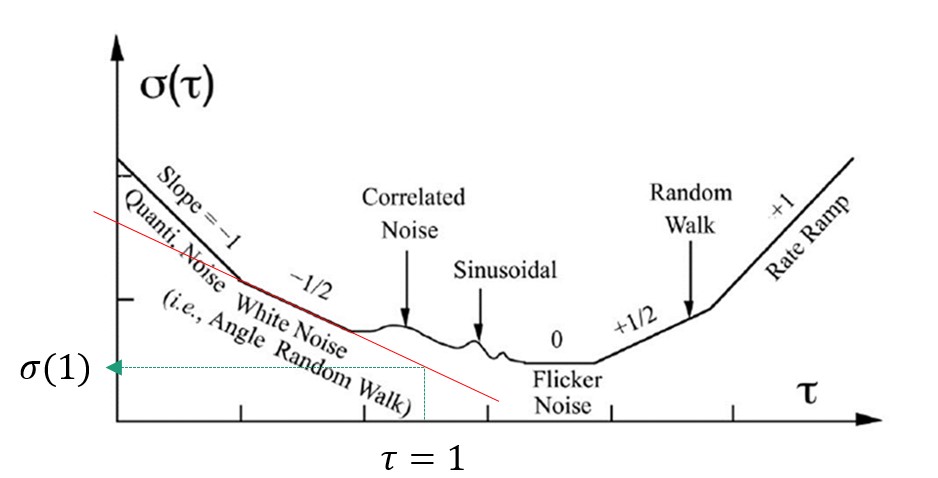}
            \vspace{-\baselineskip}
            \caption{Different noise components are prominent at different duration of AD curve and standard deviation of white Gaussian noise can be derived when $\tau=1$ ~\cite{nirmal}. } 
            \label{fig:AD}
            \vspace{-\baselineskip}
            \end{figure}

EEG signals are inherently non-Gaussian~\cite{non_Gaussian_eeg}. Hence, as Figure \ref{fig:ica_impact} suggests, the presence of Gaussian noise has little impact on the expected functionality of commonly used EEG filtering algorithms like Independent Component Analysis (ICA)~\cite{Braun_2021}.
This gives us confidence that remotely injected Gaussian noise onto EEG headsets for providing DP will not adversely affect EEG sensing accuracy.
To accurately quantify the noise introduced in EEG signals, we need a reliable method to analyze its statistical properties. Allan Deviation (AD) analysis is widely used to identify white Gaussian noise and estimate its SD in sensor noise~\cite{nirmal,Zhao2013ATA,Regression_av,visual_light_av,gps_av}. In AD analysis, the continuous time signal is divided into overlapping clusters, each with $n$ overlapped series from the original series over a duration $\tau$. The signal is then averaged over $\tau$ to generate a new timeseries that represents variations at different averaging intervals. The variance among these groups is computed as function of $\tau$, and AD is defined as the square root of this variance. As shown in Figure \ref {fig:AD}, different stochastic noise components dominate at different regions of $\tau$, allowing sub-noise categories such as white Gaussian noise to be independently profiled and studied. Furthermore, the AD curve exhibits a slope of -0.5 in the presence of white Gaussian noise, and its SD is equal to the AD when $\tau = 1$ ~\cite{nirmal,Zhao2013ATA,Regression_av}. Given these properties, we adopt AD analysis as a reliable method for quantifying the white Gaussian noise induced on EEG headsets.

\subsection{DP noise requirement in EEG based community sensing}
\begin{figure}[hbt!]\vspace{-\baselineskip}
            \centering
            \includegraphics[width=\linewidth]{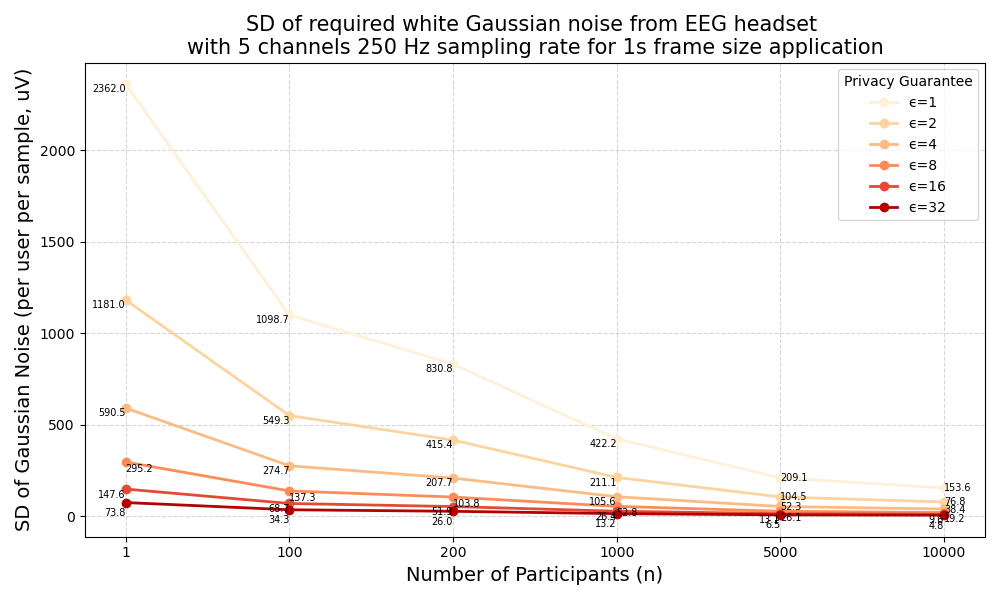}
             \vspace{-\baselineskip}
            \caption{Minimum noise's standard deviation required per user per  sample for different DP guarantees ($\epsilon$) w.r.t. a given population size.}
            \label{fig:DP_requirement}
            \vspace{-\baselineskip}
\end{figure}

To determine the minimum noise required to guarantee DP, we follow the theorems in Section \ref{Differential_privacy_theorems}. Figure \ref{fig:DP_requirement} depicts the required noise for different community sizes under a range of privacy guarantee levels. Since EEG-based community sensing focuses on classroom and factory floor environments, a reasonable target community size is 100 ~\cite{Space_Planning_Guidelines}. For 100 participants, community sensing program needs 34.33 $\mu$V of SD of white Gaussian noise per sample at the user level to facilitate DP with $\epsilon=32$ for 1 second 5-channel EEG system data sampled at 250 Hz rate. 

According to DP theorem \ref{equation:8}, SD of required DP noise per sample in 1 second data frame increases proportionally to square root of sampling rate and square root of number of used EEG channels. Hence community sensing program with 100 participants needs 21.71 $\mu$V SD of DP noise when data are arriving from 1-channel EEG system sampled at 512 Hz compared to 14.07 $\mu$V for 1-channel EEG headset with 215 Hz sampling rate.

%% file: sections/Related_work.tex
\section{ RELATED WORKS}
\subsection{DP for wearable community sensing}
In 2017, Apple introduced DP in macOS Sierra and iOS 10~\cite{apple_priv} to ensure that analyses of user behavior and usage patterns could be performed without exposing identifiable user data. Recent Exposure Notification Privacy-preserving Analytics (ENPA) introduced by Apple and Google to enable automated alerts to users with potential exposure to COVID-19 has also used DP~\cite{enpa}.  Existing deployments of DP for wearable-based community sensing are limited to smartphone-based systems. This is largely because wearables such as fitness trackers, smart watches and EEG headsets provide limited configurability, compute, and power resources~\cite{serandip}.


\subsection{DP without random noise generation}

Several recent works propose approaches for providing DP without noise generation.
Duan et al.~\cite{yitao} mathematically validated that the inherent uncertainty associated with unknown quantities of noise available in the data can be used to guarantee privacy without adding external noise, when the data is sufficiently large in a centralized server setting.
Bhaskar et al.~\cite{noiseless_microsoft} considered that the aggregated data itself has sufficient statistical noise to provide demanding DP guarantee, under the following few conditions: adversarial attackers' limited knowledge regarding the data stored in the server and that each user entry in the server is independent. Such assumptions do not hold for EEG-based community sensing programs as Shukla et al. were able to build up a correlation model between hand movements data collected from Sony smartwatches and Emotiv Epoc+ EEG data to conduct EEG attacks~\cite{correlation_eeg}.
Both~\cite{yitao} and~\cite{noiseless_microsoft} trust the centralized community sensing server where DP is guaranteed. In our work, we consider EEG based community sensing servers to be untrustworthy and ensure DP protection at the sensor hardware on the EEG headset before releasing the EEG data to the server.

Kalupahana et al.~\cite{serandip} proposed Serandip, which leverages inherent sensor noise in wearable sensors - accelerometer, barometer and temperature sensor by changing the sensor configurations (sampling rate, filter cut-off frequencies and range) to generate data with a sufficient amount of inherent noise for differential privacy. 
Based on our exploratory study on Serandip's applicability to existing EEG systems (Appendix \ref{appenix1}), the inherent sensor noise of OpenBCI EEG system at the lowest gain configuration is 5.7 $\mu$V with the notch filter on, and 38.4 $\mu$V with the notch filter off. So the OpenBCI EEG headset can satisfy the DP requirement of 34.33 $\mu$V from both inherent noise and environmental noise at a lower gain. However, environmental noise is highly unpredictable. 
Hence OpenBCI's sensor noise is insufficient to provide DP reliably, even with the notch filter turned off, and at lower gain.
The maximum achievable SD of white noise from non-configurable consumer-grade EEG headsets: Neurosky and Sichiray is 3.46 $\mu$V, which is way below the required 21.71 $\mu$V for 1-channel EEG system based community sensing program with 100 participants.
This limits Serandip's applicability to EEG-based community sensing programs.

\subsection{Privacy and security for EEG and BCI}
{\bf Privacy leakage and security attacks:}
Learning attacks are conducted to corrupt EEG analytics by perturbing EEG channel inputs~\cite{Data_analytic_attack_pertubation1, data_anlytic_perturbation2, backdoor_attack}. These noise-based perturbation attacks are performed on EEG data at different stages of the EEG data pipeline, mainly data acquisition~\cite{brain_hack,first_to_EEG_attack} and processing~\cite{noise_based_cyber_attack,backdoor_attack, SSVEP_attack}. Beltran et al. claimed that an attacker with knowledge can reduce accuracy of EEG data analytics by upto 22\% by conducting noise-based EEG attacks at the acquisition level~\cite{noise_based_cyber_attack}.
Amplitude modulated RF signals have been recently employed to conduct end-to-end perturbation attacks on EEG data at acquisition level from off-the-shelf EEG headsets~\cite{brain_hack,first_to_EEG_attack}. Xiao et al. were successful in stealing NeuroSky EEG data via a proximate attack and remote attacks while achieving 70.55\% inference accuracy~\cite{EEG_security_attack}. Hence, privacy protection for EEG data produced by EEG headsets must be provided at the data acquisition stage and data needs to be encrypted before transmission. 
\vspace{-0.2\baselineskip}

{\bf Privacy and security protection:}
Though there are multiple works regarding EEG attacks to corrupt and steal EEG data, to the best of our knowledge, no prior works targeted the privacy and security protection of EEG data. End-to-end attacking papers propose application-specific solutions like frequent change in frequencies of SSVEP control signals, response time personalization and common-mode rejection as algorithmic level countermeasures while shielding and active electrodes as design-level measures~\cite{brain_hack,first_to_EEG_attack}. We also note that Emotiv applies AES encryption on data before sharing with the server to avoid EEG data stealing attacks~\cite{EMOTIV}. For EEG-based community sensing, Differential Privacy has been proposed but not yet realized~\cite{Ienca_Fins_Jox_Jotterand_Voeneky_Andorno_Ball_Castelluccia_Chavarriaga_Chneiweiss_et, Yuste_Goering_Arcas_Bi_Carmena_Carter_Fins_Friesen_Gallant_Huggins_et}.

%% file: sections/System_Design.tex
\section{ E-MAGDIP SYSTEM DESIGN }
\label{sec:5}

\subsection{ System overview }
\begin{figure} [hbt!]\vspace{-0.5\baselineskip}
    \centering
    \includegraphics[width=0.8\linewidth]{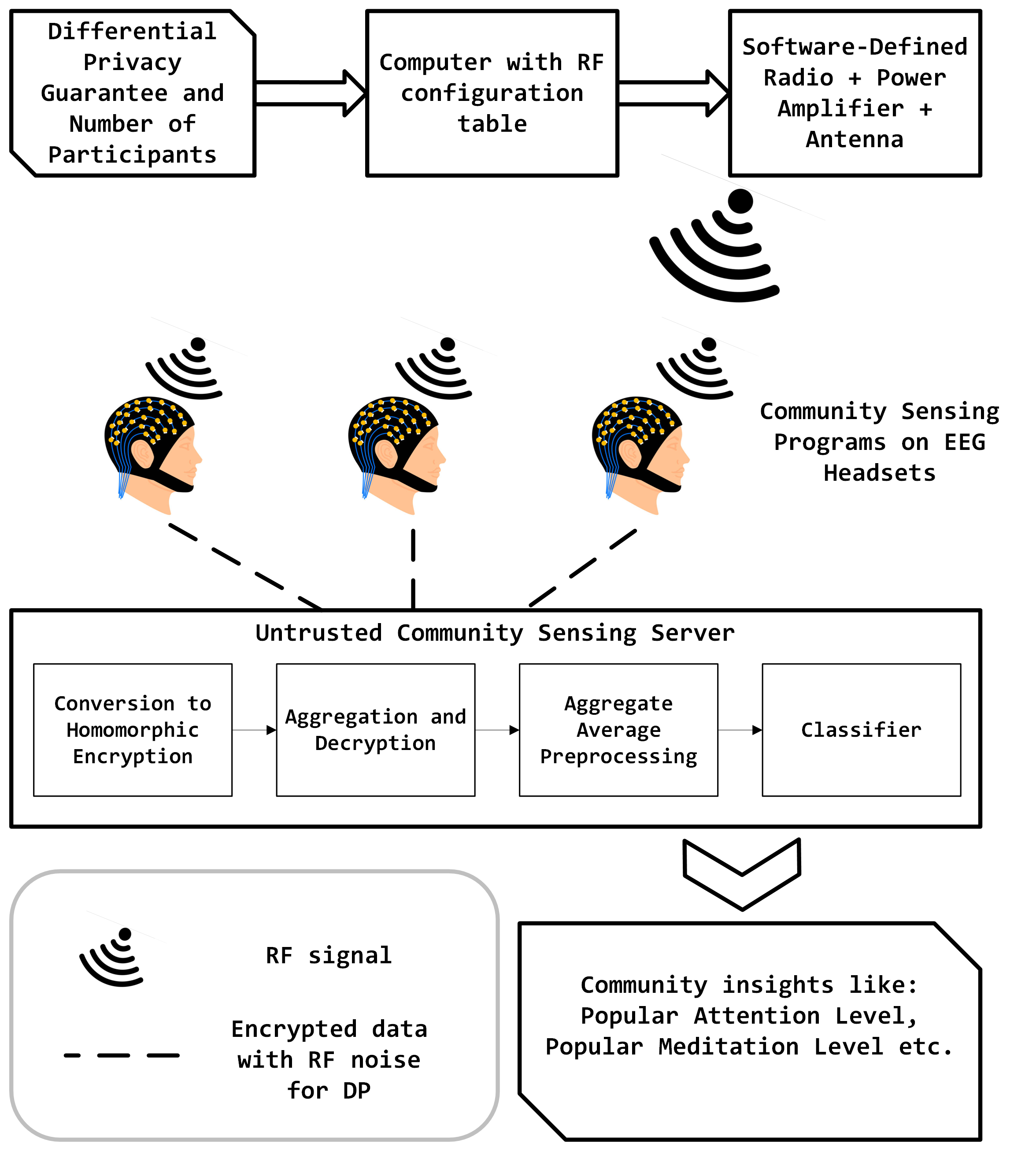}
    \caption{System overview}
    \label{fig:overview}
    \vspace{-\baselineskip}
\end{figure}

E-MagDiP consists of three main components: a DP noise transmission system using RF, EEG headsets, and an untrusted community sensing server, as illustrated in Figure \ref{fig:overview}. The DP noise transmission system determines the required SD of Gaussian noise based on the DP guarantee and the number of participants. The software-defined radio (SDR) setup, which includes a power amplifier and a transmitting antenna, is then configured to transmit the required amount of noise. Notably, E-MagDiP does not require any hardware or software modifications to the EEG headsets. EEG data is collected in AES-encrypted form by the untrusted community sensing server. To enable computations on the encrypted data, AES-encrypted signals are first converted to a homomorphic encryption format using an open-source AES-to-homomorphic conversion library ~\cite{Halevi2020DesignAI}. This conversion allows computations to be performed on encrypted data without exposing raw signals. The server then aggregates the homomorphically encrypted data, applies decryption, and computes the average EEG values across participants similar to 
other EEG-based community sensing aggregation techniques~\cite{EEG_aggregation, serandip}. Preprocessing techniques such as notch, band-pass  filtering and wavelet based denoising are applied before classification algorithms to determine community-level cognitive activities, such as finding the most common attention level or meditation state.
\subsection{ Materials}
\begin{figure}[hbt!]\vspace{-\baselineskip}
\centering
\includegraphics[width=1\linewidth]{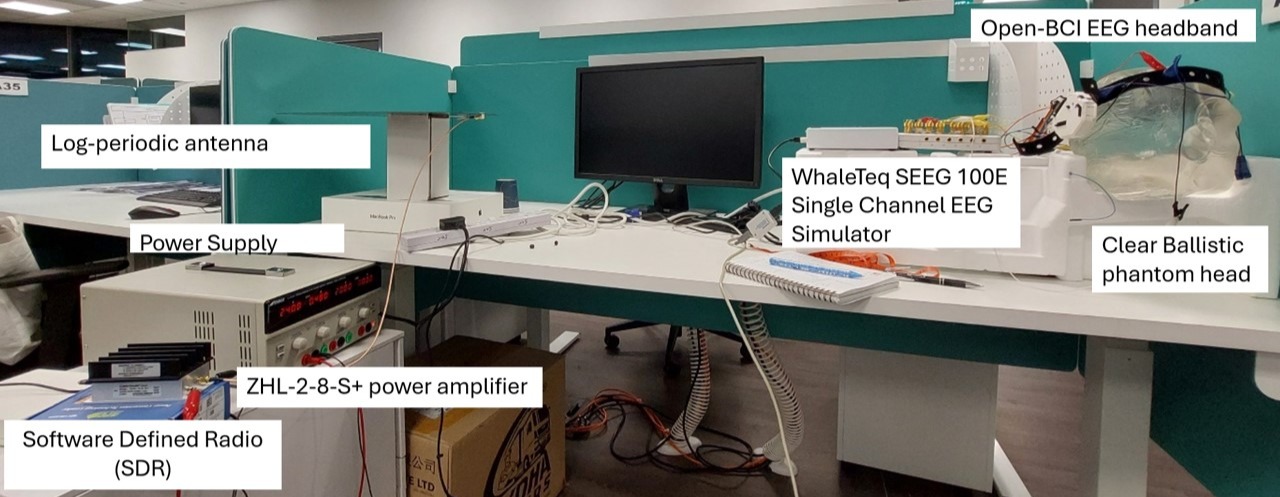}
\caption{The E-MagDiP system setup as shown with a phantom head was used during initial experiments.}
\label{fig:initial}
\vspace{-\baselineskip}
\end{figure}

Figure \ref{fig:initial} shows the various components that form E-MagDiP experimental setup, and we walk through each in detail next.

\subsubsection {EEG headsets:}
\begin{figure}[hbt!]
\centering
\includegraphics[width=0.7\linewidth]{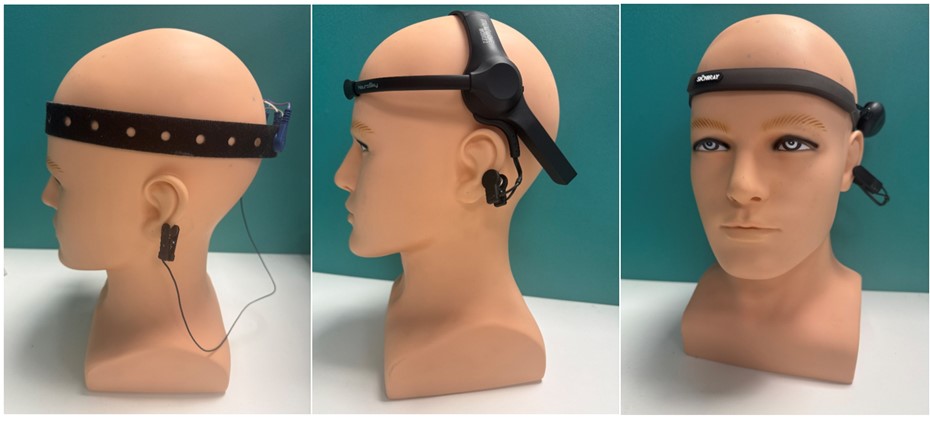}
\caption{The EEG systems: OpenBCI, Neurosky and Sichiray  used in E-MagDiP experiments.}
\label{fig:eeg_headsets}
\vspace{-\baselineskip}
\end{figure}

This section describes the EEG headsets (see Figure~\ref{fig:eeg_headsets}) used to test and validate E-MagDiP.
 
\begin{itemize}              
\item {OpenBCI EEG headband~\cite{OpenBCI_}:}
The OpenBCI EEG headband is a research-grade EEG headband compatible with all OpenBCI biosensing boards: Cyton~\cite{OpenBCI_cyton}, Ganglion~\cite{OpenBCI_ganglion}, and CytonDaisy~\cite{OpenBCI_cyton_daisy}. For E-MagDiP experiments, we interfaced the OpenBCI EEG headband with the Cyton OpenBCI biosensing board configured according to Table \ref{tab:configurations_neurosky_emagdip}. The OpenBCI GUI Impedance widget for Cyton ~\cite{OpenBCI_impedance_widget} was used to ensure proper electrode connectivity to the scalp. Data collection was performed using a Python script developed with the open-source BrainFlow library ~\cite{Brainflow}. \label{OpenBCI}
\item{ Neurosky Mind-Wave Mobile
2 EEG headset~\cite{neurosky}:} Neurosky is a consumer-grade EEG system (see Table \ref{tab:configurations_neurosky_emagdip} for configured factory settings) with a single dry electrode positioned on the forehead at the Fp1 location.  EEG data was collected using a matlab script developed by interfacing with ThinkGear connector~\cite{thinkgear}.
\item{Sichiray Taurus 2.0 Brainwave headband~\cite{sichiray}: Sichiray is another consumer grade EEG system (see Table \ref{tab:configurations_neurosky_emagdip} for configured factory settings) with single gold plated dry electrode to capture EEG signal at Fp1 location. Since the Sichiray headband also uses ThinkGear connector, the same matlab script developed for Neurosky was used for data collection. }
\end{itemize}

\begin{table}[hbt!]
            \caption{Configuration parameters for OpenBCI EEG headset and factory settings for Neurosky Mind-Wave Mobile 2  EEG headset and Sichiray Taurus 2.0 Brainwave headband used in E-MagDiP experimental setup}
             \vspace{-\baselineskip}
  \label{tab:configurations_neurosky_emagdip}
  \scalebox{0.6}{
  \begin{tabular}{ccc}
    \toprule
     \bf{EEG Headset Model} &
      \bf{Configuration Parameter } & \bf{Value}\\
      \midrule
      OpenBCI +Cyton &   Sampling Rate & 250Hz
       \\
        & Gain & 1 \\
       & Notch filter activation &OFF\\
       & No. of channels  & 5 \\   
    \midrule
      Neurosky Mind-Wave Mobile 2  &Sampling Rate & 512Hz
       \\
       &Notch filter activation &ON \\
       &No. of channels considered & 1 \\
      \midrule
      Sichiray Taurus 2.0 Brainwave 
            &Sampling Rate & 215Hz
       \\
       &Notch filter activation & ON\\
       &No. of channels considered & 1 \\ 
            
        \bottomrule
    \end{tabular}
    }
    \vspace{-\baselineskip}
\end{table}            

 
\subsubsection{Software-defined radio:}
A Software-Defined Radio  (SDR) is a RF communication system in which wireless communication components are implemented digitally using signal processing algorithms instead of analog circuitry.  In this study, SDR was programmed using the open-source radio toolkit, GNU Radio~\cite{GNU}. We used USRP-B200~\cite{Ettus} as the SDR for our evaluations. The maximum transmission power achievable with the SDR is 20 dBm; however, this decreases significantly due to signal attenuation during wireless transmission.
\vspace{-0.5\baselineskip}
\subsubsection{Directional antenna 
:}
To transmit RF signals, the  LPDA $\_$MAX log-periodic antenna~\cite{Rfspace} was used due to its directional properties. This antenna is designed to provide a gain of 4–7 dBi over a frequency range of 300–1000 MHz.
 \vspace{-0.5\baselineskip}
\subsubsection {Omni-directional antenna
:}
In environments where individuals, such as students in a classroom or workers on a factory floor, are spatially distributed, RF signals must be transmitted in multiple directions rather than a single direction. To address this, experiments were expanded to include the omni-directional ANT500 antenna~\cite{ANT500}, which supports a frequency range of 75–1000 MHz, in addition to the log-periodic antenna.
\vspace{-0.5\baselineskip}

\subsubsection {Phantom head and EEG simulator: }
\label{phantom_head_sim}
For our characterization experiments 
, we used a Phantom (10\% Joe Fit Head Gel) from Clear Ballistics~\cite{ClearBallistics} to hold the OpenBCI EEG headset. The EEG electrodes were connected to the WhaleTeq SEEG 100E Single Channel EEG Test Unit which generates EEG test signals~\cite{WhaleTeq}.
 \vspace{-0.5\baselineskip}
\subsubsection {Digital signal processing inside SDR:}   
A simple amplitude modulator was implemented in SDR using GNU Radio App's software blocks where a Gaussian noise signal with a standard deviation of 35 was used as the modulating signal. 
\vspace{-0.5\baselineskip}
\subsubsection {Power amplifier (PA):} 
We use the ZHL-2-8-S+ power amplifier from Mini-Circuits~\cite{Mini-circuits} between the SDR and antenna, to amplify the RF transmit power and induce higher Gaussian noise. Our power amplifier provides a gain of 35 dB (approximately at 24 V DC supply voltage for 10-1000 MHz frequency). 

\subsection  {RF system design and calibration}
\subsubsection{Carrier frequency:}
For optimal radio frequency transmission, the carrier frequency must be determined considering the geometric dependence of the receiving antenna. Due to this, carrier frequency must be chosen empirically or based on prior knowledge of the system. In an EEG attacking scenario, this is done by sweeping through a reasonable frequency range while monitoring the output of the target system. The expected range for the carrier frequency ($f_c$) can be determined mathematically, where it is related to the length (L) of electrode wires used in the system: $f_c$ = c/L, where c is the speed of light~\cite{brain_hack}.
Based on our measurements, the OpenBCI headset's electrode length is 56cm for the ground electrode and 1.5m for other electrodes which demand 535.7 MHz and 200 MHz carrier frequencies respectively. Since the closest ISM band to the ground electrodes is 433.05-434.79 MHz, we selected 434 MHz as the carrier frequency for the OpenBCI headband. The SD of the induced white Gaussian noise does not exhibit significant variation across different carrier frequencies. In particular, the SD measured at 434 MHz was 52.54 $\mu$V, while at 535.5 MHz it was 50.78 $\mu$V. These measurements were taken under the following conditions: the notch filter was disabled, the EEG headset amplifier gain was set to 1, the SDR gain was set to 1, and both the carrier and modulation amplitudes were set to 100. The SD of the modulating Gaussian noise was 35. As the Neurosky Mind-Wave headset and Sichiray Taurus 2.0 Brainwave headband have shorter exposed reference electrode wires, we swept frequencies and identified 757 MHz as their best matching carrier frequency.

\subsubsection{Effect of SDR and Power Amplifier on SD of Gaussian Noise:}
\begin{figure}[hbt!]\vspace{-1\baselineskip}
\centering
\includegraphics[width=\linewidth]{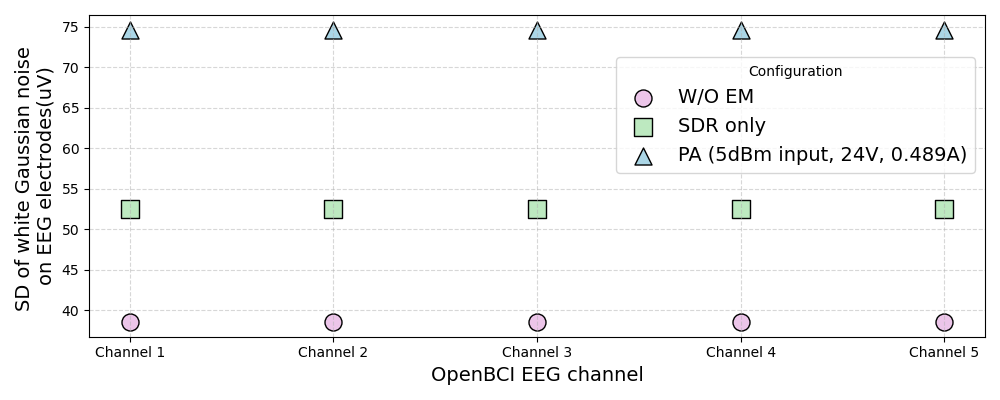}
 \vspace{-\baselineskip}
\caption{Effect of power amplifier on SD of white Gaussian noise on OpenBCI EEG headset.}
\label{fig:power_amplifier_impact}
\vspace{-1\baselineskip}
\end{figure}

Higher white gaussian noise can be induced on EEG electrodes, if the RF transmit power is higher as it increases carrier amplitude. We ran experiments with and without a power amplifier and Figure \ref{fig:power_amplifier_impact} graphs our measurements. SD of Gaussian noise increased from 30+ to 50+ as a result of the SDR RF transmission and further increased to 70+ with the power amplifier. This shows the critical impact of the PA in E-MagDiP. Further, the PA enabled the transmission of radio frequency signals over a longer distance in Section \ref{distance based DP noise}.

\subsection {RF system characterization}
\subsubsection{Transmitter and receiver distance with Differential Privacy:}
To see how well this method works over longer distances, we need to determine the potential maximum distance between the transmitter and receiver in practical deployments. For this, we consider a room that can accommodate up to 100 people. Community sensing settings would have similar footfall, schools have main halls with 17.4m x 9.8m area that can fit 100 occupants~\cite{Main_hall_hire} 
and the space norm for university auditorium and lecture theater is 1.5m² per person~\cite{Space_Planning_Guidelines}
therefore we use a room of size 17.4m x 9.8m (Area: 170.52 m² ) in our calculations. By applying Pythagoras' theorem and considering the transmitter being located at the center, the maximum distance we obtain is 9.98m between transmitter  and receiver. We place the phantom head with the OpenBCI headset and EEG simulator 
10m away from the RF setup. We observed 53.13 $\mu$V and 32.8 $\mu$V of white Gaussian noise for the directional and omnidirectional antenna respectively. In summary, the system achieved a Differential Privacy guarantee of $\epsilon = 20.68$ for the directional antenna and $\epsilon = 33.5$ for the omnidirectional antenna.
\subsubsection{Angle of incidence with Differential Privacy:}
\label{angle_section}
To explore the effect of different angles of incidence on the SD of RF based white Gaussian noise induction, we placed the phantom head equipped with OpenBCI EEG electrodes and an EEG simulator (See Section \ref{phantom_head_sim}) at varying angles relative to the directional antenna: 0$^{\circ}$, 30$^{\circ}$, 45$^{\circ}$, 60$^{\circ}$, 90$^{\circ}$, 180$^{\circ}$ and 270$^{\circ}$. The setup was maintained at a fixed distance of 1.2m between the directional antenna and the phantom head. The SDR gain was set to 1, with amplitude parameters $A_m$ = $A_c$ = 100, noise SD = 35, and no power amplifier connected. Table \ref{tab:dp_angle_of_incidence} summarizes our observations. Our results indicate that higher white Gaussian noise is induced when the EEG electrodes are exposed to the main lobe of the antenna, resulting in a stronger privacy guarantee. However, in general, the SD of noise induced on EEG electrodes and the corresponding privacy guarantee remain roughly the same across different angles of incidence, except for slight variations due to antenna radiation pattern differences.
\begin{table}[hbt!]
 \caption{Differential Privacy w.r.t. angle of incidence}
 \vspace{-\baselineskip}
  \label{tab:dp_angle_of_incidence}
  \scalebox{0.8}{
  \begin{tabular}{ccc}
    \toprule
     \bf{Angle of} & \bf{White Gaussian} &  \bf{DP Guarantee provided}\\\bf{Incidence} & \bf{Noise Induction ($\mu$V)} &\bf{ (for 100 participants)} \\
    \midrule
      0 $^{\circ}$ & 50.78 & 21.64
       \\
      30 $^{\circ}$ & 48.56 & 22.63
       \\
       45 $^{\circ}$ &47.47  & 23.15\\
       60 $^{\circ}$ &53.79  &20.43\\
       90 $^{\circ}$ & 56.27   &19.53\\
       180 $^{\circ}$ & 53.33   &20.60\\
       270 $^{\circ}$ & 45.75 & 24.02\\
        \bottomrule
    \end{tabular}
    }\vspace{-\baselineskip}
\end{table}
            
\subsubsection{RF interference on BLE communication:}
Both the intended RF exposure and OpenBCI Bluetooth communication operate in the RF domain, hence we investigated whether the RF transmission affects the Bluetooth-based EEG data transmission. We measured the packet loss in the OpenBCI GUI while varying the transmission power of the RF signal (well below the FCC threshold of 580 $\mu$W/cm$^2$) using a directional antenna and the phantom head setup. The gain of the SDR was adjusted to modulate the RF signal strength, while the distance between the transmitter and the phantom head remained constant at 1m throughout the experiment. Given the non-overlapping nature of the RF frequency (434 MHz) and Bluetooth (2.4 GHz) bands, we observed negligible packet loss across all tested RF power levels (See Table \ref{tab:RF_distance}).
\begin{table}[hbt!]

            \caption{Packet loss due to intended RF interference}
             \vspace{-\baselineskip}
  \label{tab:RF_distance}
  \scalebox{0.75}{
  \begin{tabular}{cc}
    \toprule
     \bf{RF transmitting Power (µW/cm²)} & \bf{Packet loss based on 100K packets (\%) } \\
    \midrule
      RF not applied   & 0
       \\
      557    &0\\
       279.1   &0\\
      35.1   & 0\\
       5.6 &0.001\\
       0.442   &0\\
        \bottomrule
    \end{tabular}
    }\vspace{-\baselineskip}
\end{table}

%% file: sections/Experimental_Setup.tex
\section{ EXPERIMENTAL SETUP }
\label{experimental_setup}
To evaluate E-MagDiP, we first select and implement several state-of-the-art baselines and classifier, then conduct user trials (\textit{with approval from our university's Institutional Review Board}) to evaluate the real-world impact of E-MagDiP.    
\subsection{Baselines}

\subsubsection{Software noise generation for Differential Privacy (SoftwareDP):}
Most commercial EEG headsets do not allow hardware or software modifications, but research-grade EEG systems like OpenBCI's Cyton provide modifiable firmware. To compare E-MagDiP with conventional DP techniques, we implemented a state-of-the-art statistical noise generation method based on DP~\cite{dwork} in OpenBCI’s firmware. 
This baseline allows us to assess the trade-offs in energy consumption and latency introduced by conventional DP techniques when applied directly to an EEG headset. Unlike these methods, E-MagDiP does not modify EEG hardware or software, ensuring that it does not introduce additional overheads.  

\subsubsection{Sensor noise for Differential Privacy (SensorDP):}
To explore the potential of leveraging EEG inherent sensor noise for differential privacy, 
we implemented recent work Serandip by following the paper~\cite{serandip} in OpenBCI system. 

\subsubsection{No noise added for Differential Privacy (NoDP):}
To gauge the impact of RF-induced privacy noise on EEG data accuracy, we also established a baseline 
where no additional noise, either from radio frequency signals or software-based DP was introduced. This allows us to quantify any accuracy trade-offs that result from E-MagDiP’s approach.

 \subsection{EEG classifier selection}
\label{EEG_classifier}
For evaluating the accuracy of E-MagDiP against baseline methods, we decided to use a popular EEG classifier that is widely adopted in BCI applications and compatible with OpenBCI EEG data. We selected a multi-class Steady-State Visual Evoked Potential (SSVEP) classifier, as it is used in many BCI applications~\cite{SSVEP_bci_chi,SSVEP_BCI_chi2,mobisys_SSVEP} and leverages the OpenBCI EEG headband’s electrodes, which cover the occipital region— the most sensitive area on the scalp for visual processing and effectively capturing SSVEP response. 

As we are interested in exploring E-MagDiP's ability to work with existing infrastructure, we used an existing open-source SSVEP classification model ~\cite{SSVEP_classifier} for our accuracy evaluation. For this model, pre-trained model weights were not publicly available and the model was proposed to use without user-specific training. Hence, we trained the model using the popular SSVEP dataset: BETA dataset~\cite{BETA}, adhering to the methodology described in the corresponding paper~\cite{SSVEP_classifier}.

 We first evaluated the classifier using the BETA data set's 5 channels (PO5, PO6, O1, Oz, and O2) data belonging to 70 subjects with the leave-one-out cross-validation technique. To explore the potential of usability without user-specific training, we then tested BETA’s 70 user-based trained SSVEP model on the SSVEP Benchmark (SynAmps2, Neuroscan Inc.) ~\cite{Benchmark} dataset's 30-subject data. 

\subsection{User study setup}
\begin{figure}[hbt!]
               \centering
            \includegraphics[width=1\linewidth]{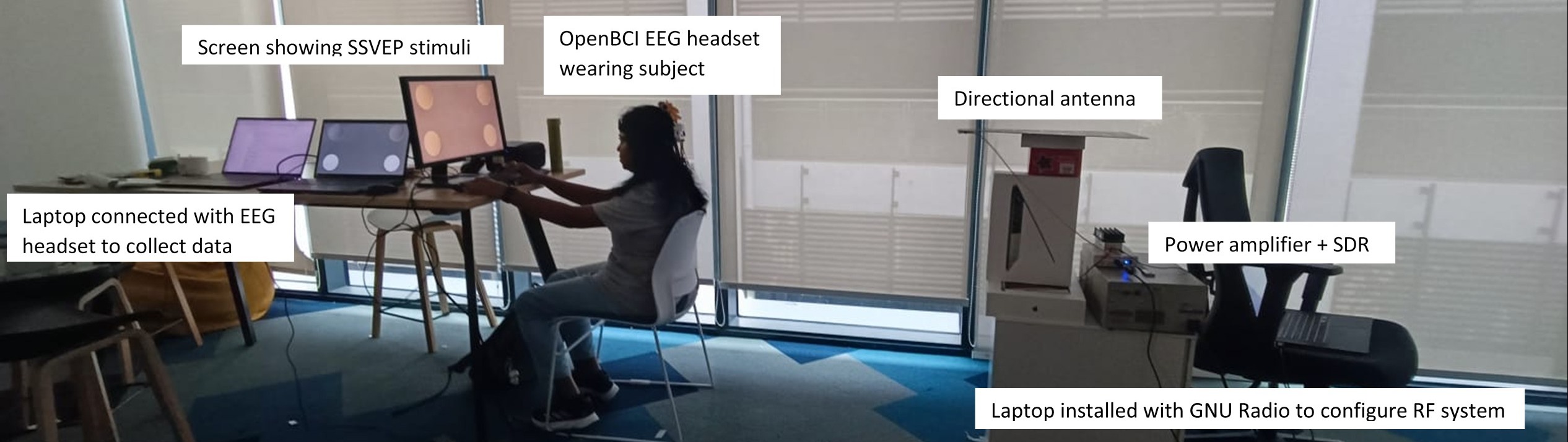}
             \vspace{-\baselineskip}
            \caption{User study setup where subject is attending to SSVEP stimuli while wearing OpenBCI EEG headset and exposed to RF signals generated by directional antenna 2 meters away to induce DP noise on EEG data.}
              \label{fig:userstudy}
        \vspace{-\baselineskip}
\end{figure}

We recruited 10 participants (5 male, 5 female) for our experiments. Each participant wore OpenBCI EEG headband (see Section \ref{OpenBCI} ) for approximately 1.5 hours. The user study was conducted in a seminar room on the university premises.
Participants were asked to stay seated while gazing at a screen with 4 Steady-State Visual Evoked Potential (SSVEP) stimuli flickering at 4 different frequencies: 11, 11.8, 14.2 and 15.8 Hz  (as shown in Figure \ref{fig:userstudy}), following a procedure similar to recent SSVEP studies ~\cite{walkingwizard}.
Participants and their headsets were exposed to noise-inducing Amplitude Modulated (SD=35,Am=Ac=100, SDR gain =0.734, power amplifier connected) RF signals (within safety regulations) at different distances (1-4m)  from the transmitter, using omnidirectional and directional antenna. Additionally, participants' 5-minute baseline data were collected while performing the same cognitive activities in a seated position without RF exposure (NoDP, SensorDP) and with software-based white Gaussian noise generation (SoftwareDP) in absence of RF exposure, for accuracy evaluation.

%% file: sections/Evaluation.tex
\section{ EVALUATION }
\label{Evaluation}

\subsection{Impact of distance on DP noise}
\label{distance based DP noise}

\begin{figure}[hbt!]
            \centering
            \includegraphics[width=\linewidth]{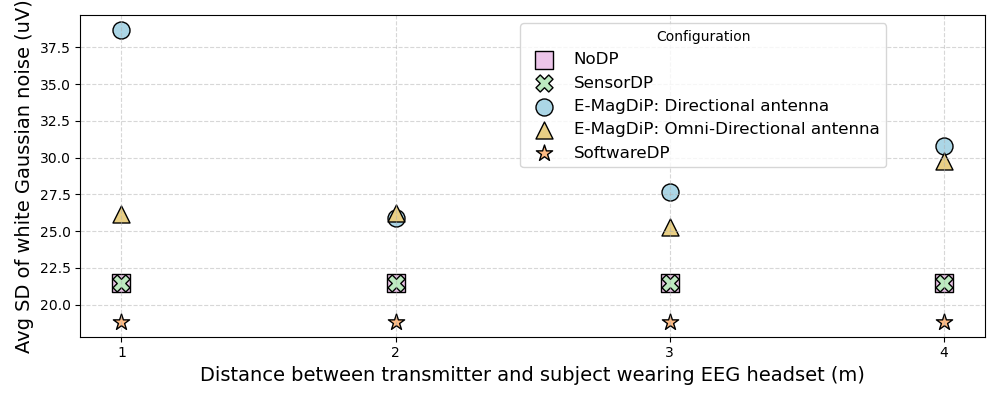} 
            \vspace{-\baselineskip}
            \caption{Average SD of the sensor noise produced on the OpenBCI headset's electrodes for 10 participants.} 
            \label{fig:OpenBCI_AVerage}
            \vspace{-\baselineskip}
\end{figure}

We first explored how RF-induced privacy noise varies with distance. This was done by varying the distance between the RF transmitting antenna and the OpenBCI EEG headset worn by participants in the user study from 1m to 4m. Figure \ref{fig:OpenBCI_AVerage} summarizes the results. The average SD of sensor noise produced by E-MagDiP on the OpenBCI EEG headset consistently showed higher noise induction with RF exposure as compared to the baselines (SoftwareDP, SensorDP and NoDP). However, we did not observe a clear trend in SD variation as a function of distance. Specifically, the SD for the 4m distance was slightly higher than that for the 2m distance, but the difference was negligible. This suggests that the RF transmitter can be rather flexibly located, easing practical deployment of E-MagDiP.

\begin{figure}[hbt!]
            \centering
            \includegraphics[width=.9\linewidth]{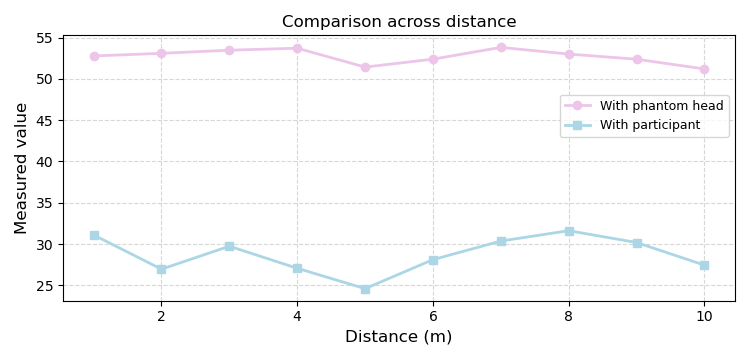} 
            \vspace{-\baselineskip}
            \caption{Average SD of the sensor noise produced on the OpenBCI headset's electrodes for 1 participant Vs phantom head while varying distance between directional antenna based RF transmitter and user/phantom head.} 
            \label{fig:distance_vs_noise}
           
\end{figure}

We further investigated the effect of distance on OpenBCI RF based noise induction by varying the distance from 1-10m between directional antenna based RF system (Am=Ac=100, SD=35, SDR gain =0.734, Power amplifier connected and fc=434 MHz) and OpenBCI EEG headset placed on a user and phantom head at 1m intervals. As the phantom head setup was connected to the EEG simulator via wires, it induced considerably higher privacy noise (Avg. SD 52.73 $\mu$V ) than that worn by a user (Avg. SD 28.72 $\mu$V ). The phantom head setup was also immune to the minor temporal movements of the human while exposing to RF, leading to min-max difference between induced noise over 1-10 m of 2.6 $\mu$V for phantom head compared to 7.02 $\mu$V for the human subject. As shown in Figure~\ref{fig:distance_vs_noise}, we did not observe any clear trend of SD of white Gaussian noise induction on both phantom head connected OpenBCI EEG electrodes and human head connected EEG electrodes while line of sight RF exposure is maintained. Hence, this suggests our 1-4m user study can be expanded to real world deployments that span larger area.

\subsection{Impact of antenna type on DP noise} 

To explore the impact of antenna type on DP white Gaussian noise induction, the user study was conducted with both directional and omnidirectional antennas. A directional antenna focuses RF transmission in a specific direction, delivering higher power along the main lobe, whereas an omnidirectional antenna transmits RF signals uniformly in all directions with lower power. For the directional transmission, we aimed the main lobe at the EEG system's reference electrode. Other electrodes could still receive RF exposure from side lobes, albeit at a lower transmission power. Consequently, the side lobe exposure is expected to induce lower amplitude white Gaussian noise on the EEG electrodes. In contrast, an omnidirectional antenna distributes RF power supply along all directions, meaning the induced noise across electrodes is expected to be higher than side lobes, but lower than the main lobe of a directional antenna. Since the total induced white Gaussian noise is an accumulation of all RF exposure, it is not immediately clear whether a directional antenna would always induce more noise than an omnidirectional one. Based on our user study with the OpenBCI EEG system (Figure \ref{fig:OpenBCI_AVerage}), the directional antenna induced significantly more white Gaussian noise at 1m distance as compared to the omnidirectional antenna. However, at distances of 2m, 3m, and 4m, the difference in SD of the induced noise between the two antenna types was minimal. These results suggest that the choice of antenna has only a minor impact on E-MagDiP.
Such flexiblity eases the practical deployment of E-MagDiP, as form factor and cost can be taken into account when selecting the antenna.
\subsection{Quantitative evaluation of privacy}

 Due to practical constraints, we limited our user study to 10 participants. The accumulated white noise in aggregated data is thus less than the targeted population of 100 participants. Based on our user study on OpenBCI EEG system, the average SD of white Gaussian noise induced from RF is 28.8 $\mu$V compared to 21.49 $\mu$V from SensorDP (Figure \ref{fig:OpenBCI_AVerage}). Hence by applying DP theorem \ref{equation:8}, the community sensing setup with 10 users with E-MagDiP provides Differential Privacy guarantee of 120.63 to the classifier output which uses 1s frame data of 5 EEG channels as inputs.
 The baseline, SensorDP provides DP guarantee of 161.66. According to DP concept, a system provides higher privacy protection if the DP guarantee value is lower~\cite{DPbook}. This implies that E-MagDiP provides stronger privacy protection than SensorDP. Since the same procedure can be expanded to 100 participants, the DP guarantee that we can provide in classroom settings with E-MagDiP can be readily improved to 38.12.

\subsection{Scalability of 10 participant user study for 100 participant community sensing} 

\begin{figure}[hbt!]
             \vspace{-\baselineskip}
            \centering
            \includegraphics[width=\linewidth]{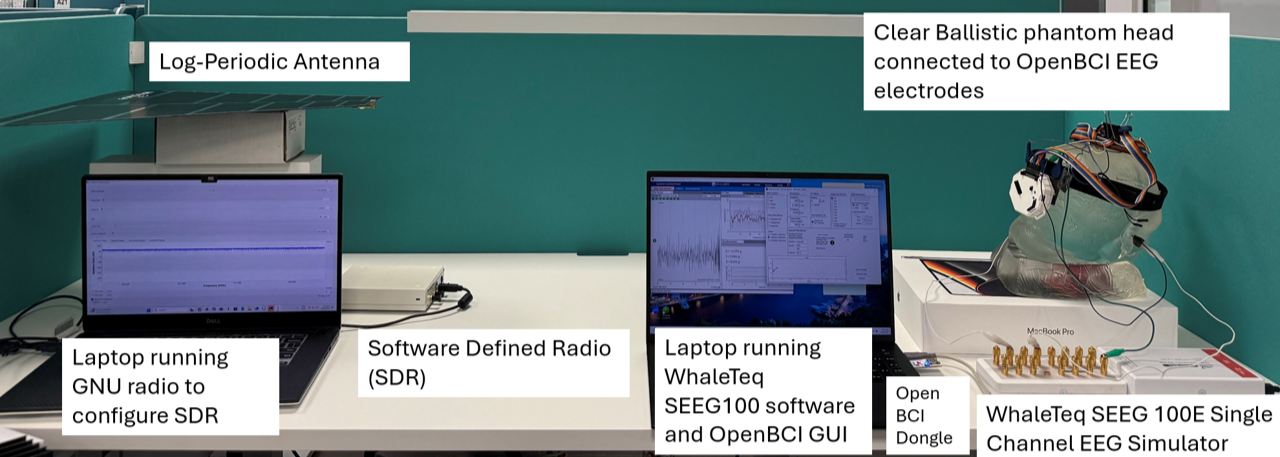} \vspace{-\baselineskip}
            \caption{OpenBCI EEG headset based scalability experiment to show the scalability of 10 participant user study for 100 participant community sensing.} 
            \label{fig:scalability setup}
            \vspace{-\baselineskip}
\end{figure}

\begin{figure}[hbt!]
            \centering
            \includegraphics[width=\linewidth]{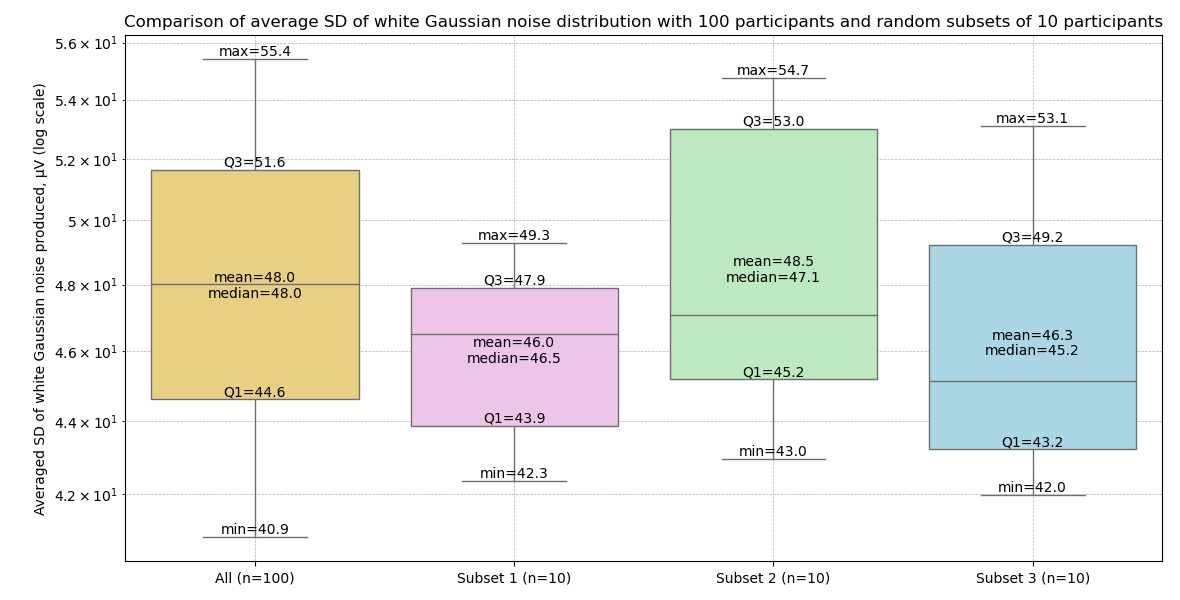}
             \vspace{-\baselineskip}
            \caption{Distribution of average SD of white noise produced by participants in 100 total participation compared to 10 participants.  }
            \label{fig:Scaling_summary}
            \vspace{-\baselineskip}
\end{figure}

To explore the potential of using 10 participant based data from user study to represent 100 participant based real-world EEG based community sensing, we used the experimental setup shown in Figure \ref{fig:scalability setup}. OpenBCI EEG electrodes were worn on the phantom head while connected to Whaleteq SEEG100 Single channel EEG simulator.  WhaleTeq SEEG100 software running on laptop configures the EEG simulator to generate EEG signals. For EEG data recording, OpenBCI GUI communicated with OpenBCI Cyton board via OpenBCI dongle. In parallel, SDR generated AM signal based on configurations received from GNU Radio software running on another laptop. We directly interfaced SDR with a log-periodic directional antenna, without a Power Amplifier, as this setting suffices to meet DP noise requirement with EEG simulator. We then fed 100 EEG records (Oz position) related to 100 participants from the EEG Motor Movement/Imagery Dataset~\cite{Physionet} to the simulator for 5 minutes. RF transmission was conducted with the same settings (Am=Ac=100, SD=35, SDR gain =1 and fc=434 MHz) as configured in section \ref{angle_section}. We collected 5 minutes of data from each participants and measured average SD of white gaussian noise induced. 
To represent the presence of 10 participants in user study, we randomly selected 3 subsets of 10 participants' average SD of white noise from 100 participants. The observed distribution of SD of induced white Gaussian noise is shown in Figure \ref{fig:Scaling_summary}. The average differences of median and mean between two distributions: 100 participants and 10 participants  are 1.74 $\mu$V and 1.4 $\mu$V respectively. Such a small difference indicates that our earlier 10-participant user study data and insights can be considered as representative of a real-world EEG community sensing program with 100 participants.

\subsection{Accuracy evaluation }
\subsubsection{Dataset based:}
We achieved 79.17\% accuracy on average (leave-one-out cross-validation accuracy) for the SSVEP classifier trained on BETA dataset's 5 channels (PO5, PO6, O1, Oz, and O2) of 70 subjects for 4 SSVEP frequency classes (11, 11.8, 14.2 and 15.8 Hz) where the probability of random guess is 25\%.  Average accuracy dropped to 68\% when testing 70 subject-based BETA dataset-trained models on the Benchmark dataset's 30 users. 
We use this 70 subject-based BETA dataset-trained SSVEP classifier model for E-MagDiP's user study accuracy evaluation. 
     

\subsubsection{User study based:}
\begin{table}[hbt!]
\vspace{-\baselineskip}
 \caption{User study based accuracy}
  \vspace{-\baselineskip}
  \label{tab:user_study_acc}
  \scalebox{0.85}{
  \begin{tabular}{cccccc}
    \toprule
     \bf{EEG Headset} & \bf{E-MagDiP} &  \bf{NoDP} & \bf{SoftwareDP} & \bf{SensorDP}\\

    \midrule
     OpenBCI& 69.4\% & 72.5\%& 68.2\% &  72.5\%
       \\
       \bottomrule
    \end{tabular}
    }\vspace{-\baselineskip}
\end{table}
In our user study, we aggregated the EEG data collected from all 10 users at the server, and applied pre-processing to the aggregated data (wavelet: Sym4, Bayes denoising method, soft threshold ~\cite{wavelet_based_pre_processing_algo}). 
Table \ref{tab:user_study_acc} shows E-MagDiP achieving 69.4 \% 
accuracy for 4 class SSVEP classification for E-MagDiP, 
compared to 68.2\% for SoftwareDP, and 72.5\% accuracies for the two baselines that do not meet DP requirements (NoDP and SensorDP). 
This 3.1\% drop in accuracy of E-MagDiP to support privacy is acceptable for practical realization of differential privacy in community sensing programs due to its privacy protection benefits. Further we can configure E-MagDiP's RF system to emit RF with a modulating signal, generating white noise with lower standard deviation than 35, to trade off the privacy guarantee for better accuracy. 
More participants can also be recruited to lower the amount of privacy noise needed, since privacy noise and square root of the number of participants are inversely proportional (Section \ref{DDP_theory} DDP theory).   
\subsection{E-MagDiP's compatibility with different EEG headsets}
\label{notch_other}

\begin{figure}[hbt!]
    \centering
    \includegraphics[width=\linewidth]{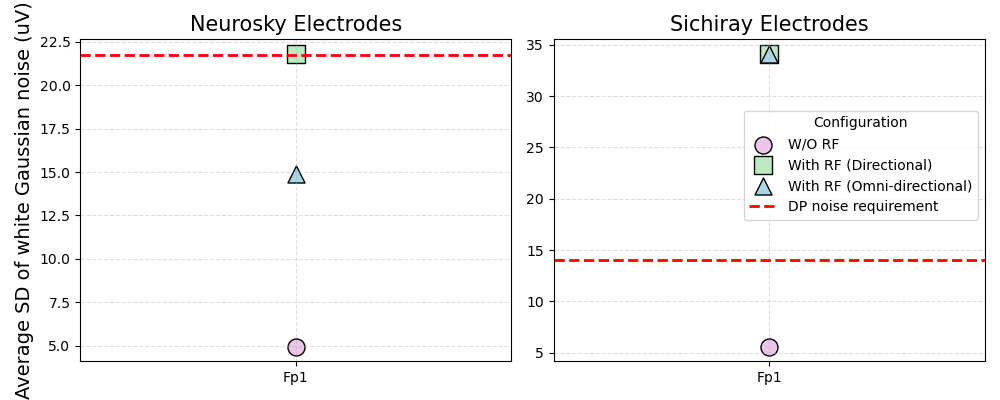} 
    \vspace{-\baselineskip}
    \caption{Average SD of the white noise induced on Neurosky Mind-Wave and Sichiray Taurus 2.0 EEG electrodes based on 10 participants.}
    \label{fig:neurosky_sichiray_user-study}
     \vspace{-1\baselineskip}
\end{figure}

We explored the potential of RF to induce DP noise on off-the-shelf consumer-grade EEG headsets: Neurosky Mind-Wave Mobile 2 EEG headset and Sichiray Taurus 2.0 Brainwave headband and the impact on EEG headset's functionality on 10 participants.
The AM signal of the white Gaussian signal (SD=35,Am=Ac=1000, SDR gain=0.734, power amplifier connected) was transmitted from the RF system with carrier frequency 754MHz (within safety regulations) at a distance of 1.2m from the transmitter. The average SD of induced white Gaussian noise on Neurosky and Sichiray EEG electrodes were 21.8 $\mu$V and 34.1 $\mu$V with directional and 14.86 $\mu$V and 34.08 $\mu$V with omni-directional antenna (see Figure \ref{fig:neurosky_sichiray_user-study}). Having passive notch filter with fixed notch depth enabled RF to induce white noise on EEG, but with lower SD. 
However with the different sampling rate and just a single electrode, the SD of required white Gaussian noise is no longer 34.33 $\mu$V. Based on DDP theories detailed in section \ref{DDP_theory}, 
the SD of required white Gaussian noise for 1s EEG data from Neurosky is 21.71 $\mu$V and 14.07 $\mu$V for Sichiray. Hence  Sichiray can facilitate DP with guarantee $\epsilon$ =32 for community sensing programs with 100 participants with both directional and omni directional antennas, while Neurosky is only able to provide the same privacy guarantee with directional antenna. But Neurosky can still provide DP for 100 with lower privacy guarantee $\epsilon=47$ with omni-directional antenna. 

When exposed to RF signals, 2-minute eye blinks can be detected perfectly across all 10 participants. 
Since blink detection is used as a cognitive indicator to assess fatigue~\cite{Blink_fatigue_chi}, mental workload~\cite{blink_rate_cognitive_load,mobisys_blink} and task difficulty~\cite{blink_cog_chi}, this provides some evidence that E-MagDiP does not adversely impact the functionality of consumer grade EEG devices.

\subsection{Simultaneous multi-user study}
\begin{figure}[hbt!]
            \centering
            \includegraphics[width=\linewidth]{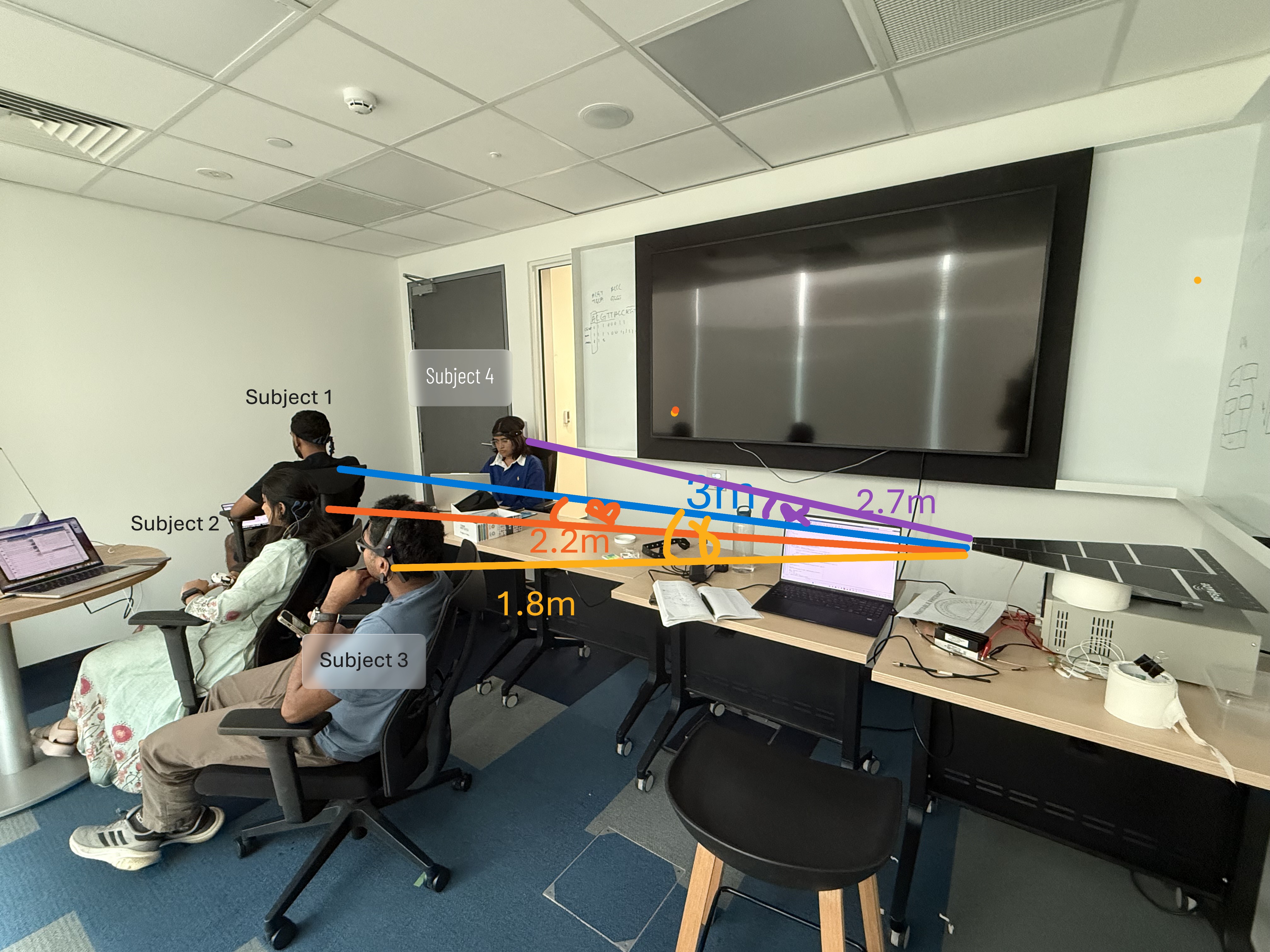} 
            \vspace{-\baselineskip}
            \caption{Multi-user experimental setup where $\alpha$ = $\beta$ =20$^{\circ}$ and $\gamma$ = 30$^{\circ}$ } 
            \label{fig:multi-user_setup}
           
\end{figure}

\begin{figure}[hbt!]
    \centering
    \includegraphics[width=0.9\linewidth]{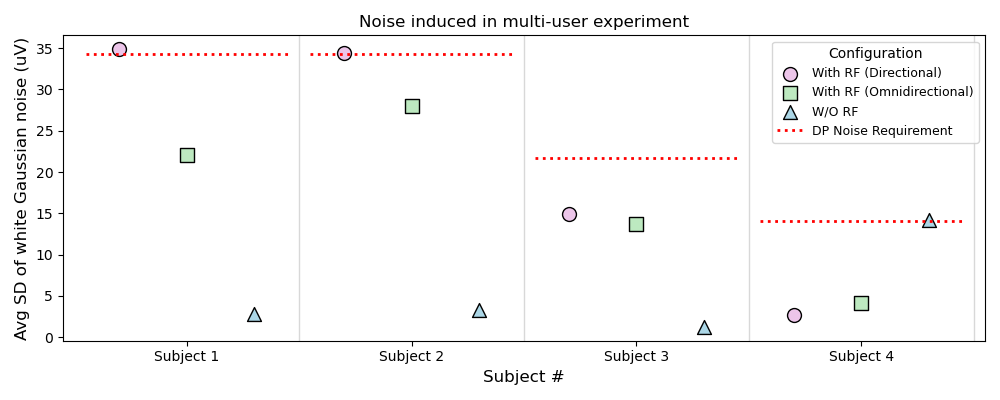} 
    \vspace{-\baselineskip}
    \caption{Average SD of the white noise induced on OpenBCI, Neurosky and Sichiray  EEG electrodes in simultaneous multi-user experimental setting.}
    
    \label{fig:simultaneous_user_study}
    
\end{figure}

To explore our framework's ability to provide DP in a classroom-like setting with multiple EEG headsets, we conducted a user study with 4 participants (Subjects 1 and 2 wearing OpenBCI, Subject 3 wearing Neurosky EEG system and Subject 4 wearing Sichiray EEG system), exposed simultaneously to RF emitting from a single RF transmitter as shown in Figure~\ref{fig:multi-user_setup}. Subject 1 maintained line-of-sight and 3m distance with RF setup compared to 20$^{\circ}$ angle and 2.2m with Subject 2, 30$^{\circ}$ angle and 1.8m with Subject 3 and 20$^{\circ}$ angle and 2.7m with Subject 4. As shown in Figure~\ref{fig:multi-user_setup}, subject 4 was located closer to many obstacles like walls, TV screen on purpose, to explore the impact from multi-path propagation. We used GNU radio's Python snippet to change carrier frequency of RF setup dynamically to support varying carrier frequency (434 MHz and 757 MHz, carrier frequency switching period = 2s )  required for different EEG headsets. The AM signal of the white Gaussian signal (SD=35,Am=Ac=1000, SDR gain=0.734,
power amplifier connected) was transmitted from the RF system with both directional and omni-directional antenna connected.
The average SD of white Gaussian noise induced on OpenBCI electrodes are 34.87$\mu$V and 34.37$\mu$V for Subjects 1 and 2 with directional antenna and it exceeds the DP noise requirement of 34.33 $\mu$V. Subject 1 and 2's OpenBCI  (with omni-directional antenna) and Subject 3's Neurosky (with both directional and omni-directional) EEG electrodes have higher noise induction, but the required DP noise requirement is not met (See Figure ~\ref{fig:simultaneous_user_study}). Since Subject 4 is in an environment with obstacles, its DP noise induction shows destructive property. 

We also evaluated correlation of Subject 1's and 2's OpenBCI EEG data with directional antenna based RF exposure, omni-directional antenna based RF exposure and without RF. Since magnitude of correlation under three conditions are within 0.3-0.4 (as shown in Table ~\ref{tab:correlation_analysis}), we can conclude that E-MagDiP based DP induction is independent despite having the same RF source. 

\begin{table}[hbt!]

 \caption{Correlation analysis of OpenBCI EEG data from 2-simultaneous users}
  \vspace{-\baselineskip}
  \label{tab:correlation_analysis}
  \scalebox{0.85}{
  \begin{tabular}{cccccc}
    \toprule
     \bf{E-MagDiP(Directional)} & \bf{E-MagDiP (Omni-directional)} &  \bf{NoDP} \\

    \midrule
    -0.31598  & 0.38927 & 0.3654
       \\
       \bottomrule
    \end{tabular}
    }\vspace{-\baselineskip}
\end{table}

\subsection{Impact on other sensors}

We verified that E-MagDiP maintains cross-sensor isolation by monitoring the OpenBCI headset's onboard accelerometer. Measurements showed that norm acceleration fluctuated minimally (0.9145g – 1.0660g) regardless of whether RF was exposed or not. This confirms our design hypothesis: by confining noise induction to non-shared EEG reference electrodes, we can successfully protect neural privacy without corrupting auxiliary data streams.

%% file: sections/Discussion.tex
\section{ DISCUSSION }
\begin{figure}[hbt!]
    \centering
    \includegraphics[width=0.9\linewidth]{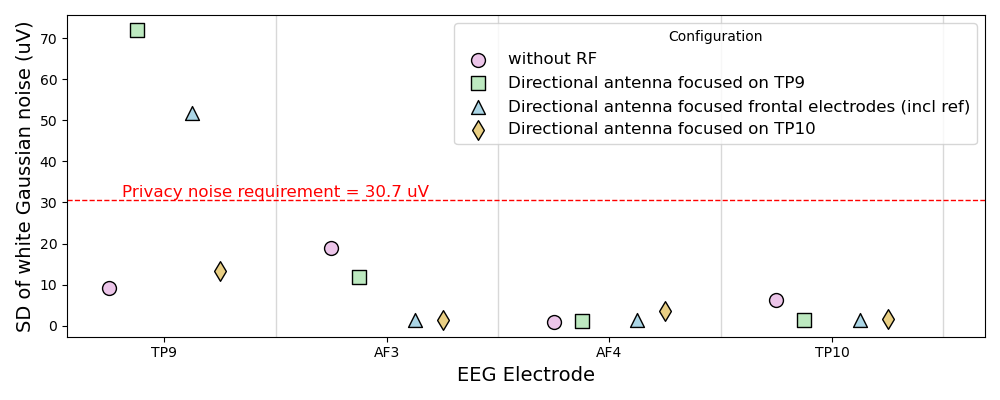} 
    \vspace{-\baselineskip}
    \caption{Average SD of the white noise induced on  non-wire exposed Muse 02-Gen 1 EEG electrodes based on one participant under different direction of exposure for 2 minutes each.}
    \label{fig:muse2}
     \vspace{-1\baselineskip}
\end{figure}

\textbf{Applicability to EEG headsets without exposed wires:}
E-MagDiP's current applicability is limited to EEG headsets with exposed wires as carrier frequency is determined based on exposed wire length. But we could induce above 70 $\mu$V SD of white noise on TP9 EEG electrode data from Muse 02-Gen1 EEG headset while a participant is wearing the headset (see Figure \ref{fig:muse2}). Here the directional antenna was focused on TP9 electrode with 1.2m distance away and transmitting white Gaussian signal (SD=35,Am=Ac =100, SDR gain =0.734, fc=500 MHz~\cite{brain_hack}, power amplifier connected).  Since the Muse 02 EEG headset does not have any exposed wires and uses flex circuitry on plastic substrates to connect the electrodes to the rest of the circuitry, RF is capable of inducing DP noise on flex circuit based EEG systems. We could also observe higher SD of noise with Muse S's TP9 electrode: 15.5 $\mu$V and Emotiv Insight-5 EEG headsets's TP8: 10.1 $\mu$V compared to other electrodes: 2-8 $\mu$V when directional antenna focused on reference electrodes (frontal for Muse S and behind the ear for Emotiv). But it's way below the privacy noise requirement of 30.7 $\mu$V (for Muse S) and 24.27$\mu$V (for Emotive Insight-5). Hence, an in-depth study is needed before a full solution can be achieved.

\textbf{Level of E-MagDiP's compatibility with notch filters:}
EEG systems use notch filters to filter out powerline interference on EEG data. Implementations of notch filter vary from simple passive filter to software-level complex adaptive filters. Hence its ability to attenuate and the attenuation level applied on noise spectrum are diverse.
Based on our experiments, we also note that 
 while E-MagDiP is effective for both Neurosky and Sichiray EEG systems (see Section \ref{notch_other} ) which have a simple passive notch filter, the complex software notch filters of OpenBCI 
 can disturb E-MagDiP's functionality (see Section \ref{OpenBCI_notch} ) as attenuation level at notch frequency dynamically changes based on signal characteristics. As future work, we plan to conduct more detailed studies on the impact of alternative notch filters on E-MagDiP. 

\begin{figure}[htb]
    \centering
    \includegraphics[width=0.8\linewidth]{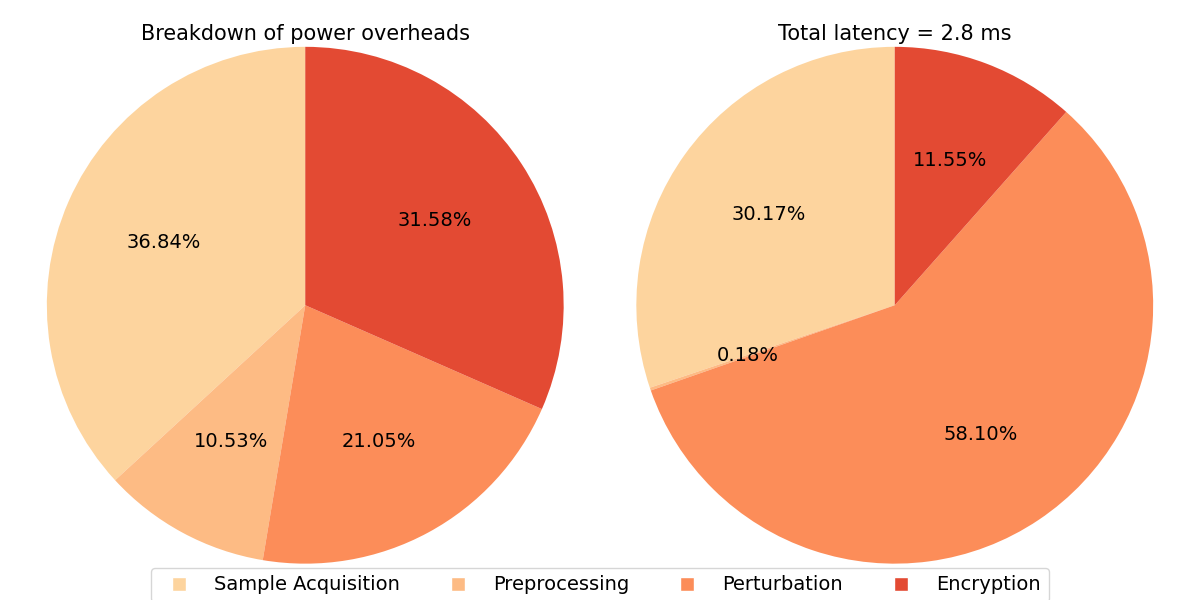}
    \caption{Measured (left) power overheads (right) latency of SoftwareDP for a sample with 8 channels of EEG data respectively.}
    \label{fig:Latency_Power}
    \vspace{-1\baselineskip}
\end{figure}

\textbf{Potential power/latency benefits over SoftwareDP:}
We implemented \textit{SoftwareDP} and encryption \cite{tiny-AES} in the firmware of the OpenBCI Cyton board. Based on our energy consumption profiling (Figure \ref{fig:Latency_Power}) using the Monsoon power monitor \cite{Monsoon_Power_Monitor}, the SoftwareDP perturbation accounts for 21.05\% of total computational energy overheads.
For fine-grained latency measurements of data acquisition and processing steps, we used a PIC32 breakout board \cite{chip_kit}, since OpenBCI has a PIC32 \cite{microchip_pic32} as its MCU. As shown in the latency profile (Figure \ref{fig:Latency_Power}), SoftwareDP perturbation contributes to 58.1\% of total computation latency.
E-MagDiP requires neither software nor hardware modifications, therefore it offers potential benefits by eliminating power consumption and latency due to perturbation as compared to SoftwareDP. Reduction in latency associated with perturbation could provide benefits to on-body devices that have multiple computational steps, like preprocessing and inference.

Additionally, when the sampling rate is 250Hz, with a 4ms sampling period, latency due to SoftwareDP can be hidden within the sampling period. 
But some EEG systems operate at higher sampling frequencies like 500Hz ~\cite{neurosky,Smarting_wireless,X,mindtooth,enobio} and 1000Hz~\cite{cyton_16kHz,bittium}, with sampling periods of 2 ms and 1 ms. SoftwareDP's computation latency exceeds these sampling periods, so, it can no longer be hidden.

Since E-MagDiP does not perform any computation at the device level, it has the potential to provide upto 1.2X computation energy savings, and 2.4x speedup in computation latency\footnote{Speedup factor at 500Hz is computed as 
\(\tfrac{2.797~\text{ms}}{1.172~\text{ms}} \approx 2.40\times\). 
Here, 2.797~ms is the total latency with perturbation, and 1.172~ms is the total latency without perturbation.}
 at 500Hz sampling rate on the Cyton board.  

%% file: sections/limitations.tex
\section{LIMITATIONS}
\textbf{Health concerns:}
Based on our RF-induced health impacts literature review, studies that highlighted negative impacts were either theoretical in nature, rat-based experiments or transmitted RF at a much higher level above the standard~\cite{wifi_no_threat,wifi_no_impact2,wifi_no_impact3}. 
Besides, the studies indicated that the Maximum
Transmission Power Density (MTPD) of any microwave frequency range (300 MHz-30 GHz) RF signal should be below the acceptable FCC threshold
580 $\mu$W/cm²~\cite{FCC}. The computed MTPD of RF signals used in E-MagDiP is 9.7 $\mu$W/cm² which is 60 times lower than the FCC threshold. Hence, harmful thermal effects will not occur due to RF exposure from E-MagDiP. 

\textbf{Regulatory concerns:}
To avoid disruptions with existing RF systems, we selected E-MagDiP's carrier frequency (for OpenBCI) as 434 MHz as it falls within the closest ISM band (433.05-434.79 MHz)~\cite{spectrum_handbook} 
for 535.7 MHz. But we used 754 MHz as the carrier frequency for E-MagDiP realization in Neurosky and Sichiray EEG systems. Since 754 MHz does not fall within the ISM band, it will be necessary to investigate a frequency within the ISM Band such as 902-928MHz before E-MagDip can be deployed commercially.

%% file: sections/conclusion.tex
\section{CONCLUSION}

This paper presents E-MagDiP, which leverages RF signals to perturb EEG data to provide differential privacy instead of attacking. This is achieved by configuring the standard deviation of the white Gaussian modulating signal according to differential privacy requirements. Extensive experimental results demonstrate E-MagDiP's ability to work with commercial EEG headsets without requiring any modification, providing 38.12 DP guarantee for community sensing with 100 participants with a slight accuracy drop of 3.1\%.


%% file: sections/Serandip_applicability.tex
\section{ Exploration of Serandip's applicability to EEG systems }
\label{appenix1}
 
\begin{figure}[hbt!]
            \centering
              
    \begin{subfigure}[t]{0.48\linewidth}
        \centering
        \includegraphics[width=1\linewidth]{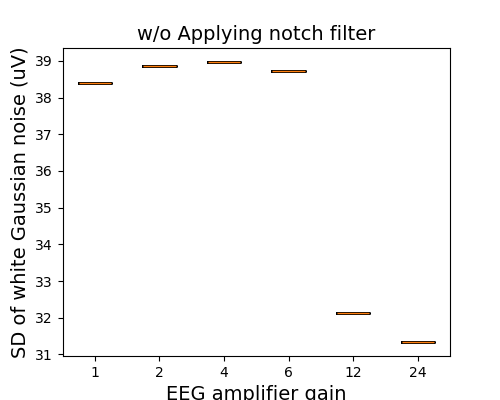}
        \label{fig:w_o_notch}
    \end{subfigure}
    \begin{subfigure}[t]{0.48\linewidth}
        \centering
        \includegraphics[width=1\linewidth]{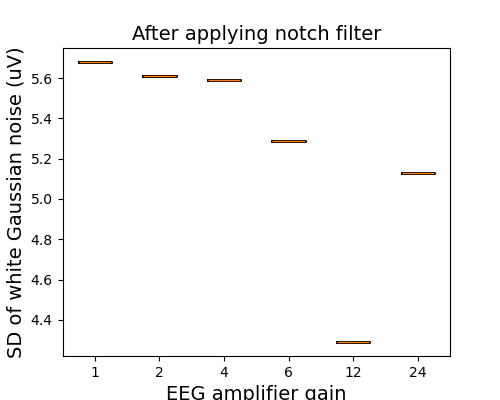}
        \label{fig:applying_notch}
    \end{subfigure}
    \vspace{-1\baselineskip}

            \caption{Serandip's noise standard deviation as measured on OpenBCI EEG electrodes+Cyton Board sampled at 250 Hz under different gain settings and with (right) /without(left) applying 50Hz notch filter}
            \label{fig:OpenBCI}
            \vspace{-\baselineskip}
  \end{figure}

To measure the effectiveness of Serandip in commercial EEG headsets, we configured the OpenBCI Cyton EEG system as shown in 
(Table~\ref{tab:configurations_inherent_sensor_noise_OpenBCI}). We then collected 5 minutes of data from the OpenBCI electrodes (connected to an EEG simulator) under each configuration, and applied the  AD algorithm to quantify the SD of white Gaussian noise introduced by the electrodes. Results showed that lower gain settings yield higher apparent noise levels, and that disabling notch filtering further increases noise due to unmitigated environmental interference (Figure~\ref{fig:OpenBCI}). In practice, EEG applications commonly apply notch filters to suppress 50/60 Hz powerline noise. Due to the reduction of powerline interference, white Gaussian nature which demands the same amplitude signal throughout all frequencies is disturbed. 
\label{OpenBCI_notch}


 \begin{table}[ht]
            \caption{Configuration parameters for OpenBCI EEG headset used in Serandip experimental setup}
   \vspace{-\baselineskip}
\label{tab:configurations_inherent_sensor_noise_OpenBCI}
  \scalebox{0.8}{
  \begin{tabular}{cc}
    \toprule
      \bf{Configuration Parameter } & \bf{Value}\\
    \midrule
      Sampling Rate & 250Hz
       \\
        Gain & 1,2,4,6,12,24 \\
       Notch filter activation &ON and OFF\\
       No. of channels  & 1 \\
            
        \bottomrule
    \end{tabular}
    }\vspace{-\baselineskip}
\end{table}

While deactivating notch filtering, the OpenBCI EEG headset can facilitate this DP requirement from both inherent noise and environmental noise at a lower gain. However, environmental noise is highly unpredictable. 
Hence OpenBCI's sensor noise is insufficient to provide DP reliably, despite turning off the notch filter.

We then expanded our Serandip experiments to two consumer-grade EEG headsets : Neurosky Mind-Wave Mobile 2 EEG headset and Sichiray Taurus 2.0 Brainwave headband. As these EEG systems do not support any device-level modifications, we had to limit our experiments to the default factory settings listed in Table~\ref{tab:configurations_neurosky}.
 \begin{table}[hbt!]
            \caption{Factory settings for Neurosky Mind-Wave Mobile 2  EEG headset and Sichiray Taurus 2.0 Brainwave headband used in Serandip experimental setup}
             \vspace{-\baselineskip}
  \label{tab:configurations_neurosky}
  \scalebox{0.6}{
  \begin{tabular}{ccc}
    \toprule
     \bf{EEG Headset Model} &
      \bf{Configuration Parameter } & \bf{Value}\\
    \midrule
      Neurosky Mind-Wave Mobile 2  &Sampling Rate & 512Hz
       \\
       &Notch filter activation &ON \\
       &No. of channels considered & 1 \\
      \midrule
      Sichiray Taurus 2.0 Brainwave 
            &Sampling Rate & 215Hz
       \\
       &Notch filter activation & ON\\
       &No. of channels considered & 1 \\ 
            
        \bottomrule
    \end{tabular}
    }
    \vspace{-\baselineskip}
\end{table}

To evaluate Serandip, we connected the electrodes of the Neurosky Mind-Wave Mobile 2  and Sichiray Taurus 2.0 EEG headsets to an EEG simulator, collected data and measured the inherent white Gaussian noise using  Matlab. Despite different factory settings, the maximum achievable SD of white noise from consumer-grade EEG headsets is 3.46 $\mu$V, which is way below the required 21.71 $\mu$V for 1-channel EEG system based community sensing program with 100 participants.